\documentclass[10pt,a4paper,twoside]{article}

\usepackage{a4wide}

\usepackage[utf8]{inputenc} 
\usepackage[TS1,T1]{fontenc}

\makeatletter

\@addtoreset{equation}{section}
\makeatother

\usepackage[french,english]{babel}

\usepackage{amssymb, amsmath,subeqnarray}

\usepackage{mathrsfs}

\newcommand{\be}{\begin{equation}}
\newcommand{\ee}{\end{equation}}
\newcommand{\bea}{\begin{eqnarray}}
\newcommand{\eea}{\end{eqnarray}}

\newcommand{\betaeff}{{\beta_\star}}
\newcommand{\C}{{\cal C}}
\newcommand{\deltaexch}{\delta_\text{\mdseries exch}}
\newcommand{\Deltaexch}{\Delta_\text{\mdseries exch}}
\newcommand{\Deltaexchexcess}{\Delta_\text{\mdseries exch}^{\text{\mdseries excs},\beta_0}}
\newcommand{\Deltaexchexcessgen}{\Delta_\text{\mdseries exch}^{\text{\mdseries excs},\{F_i^0\}}}
\newcommand{\En}{{\cal E}}
\newcommand{\Escale}{\Delta e}
\newcommand{\Esp}[1]{\langle#1\rangle}
\newcommand{\F}{\mathbb{F}}
\newcommand{\Heat}{{\cal Q}}
\newcommand{\Heatm}{q}
\newcommand{\Hist}{{{\cal H}ist}}
\newcommand{\kB}{k_{\scriptscriptstyle B}}
\newcommand{\N}{{\cal N}}
\newcommand{\Prob}{P}
\newcommand{\Probcan}{P_\text{\mdseries can}}
\newcommand{\Probeq}{P_\text{\mdseries eq}}
\newcommand{\Probmc}{P_\text{\mdseries mc}}
\newcommand{\Probst}{P_\text{\mdseries st}}
\newcommand{\Probdist}{\Pi}
\newcommand{\syst}{{\cal S}}
\newcommand{\SB}{S^{\scriptscriptstyle B}}
\newcommand{\STH}{S^{\scriptscriptstyle TH}}
\newcommand{\TimeRev}{\mathbb{T}} 
\newcommand{\Trans}{\mathbb{W}}
\newcommand{\Udisc}{{\cal T}}

\title{\textbf{
Local Detailed Balance :  A Microscopic Derivation}
}
\author{
M. Bauer\\
 Institut de Physique Théorique de Saclay\footnote{CEA/DSM/IPhT, Unité de recherche associée au CNRS}, CEA Saclay
\\ F-91191 Gif-sur-Yvette Cedex, France
 \\ \vspace{3mm}\\
 F. Cornu\\ Laboratoire de Physique Théorique,  UMR 8627 du CNRS\\
Université Paris-Sud, Bât. 210
\\ F-91405 Orsay, France
 }

\date{October 16, 2014}

\begin{document}
\maketitle

\begin{abstract}

  Thermal contact  is the archetype of non-equilibrium
  processes driven by constant non-equilibrium constraints when the latter are enforced by
  reservoirs  exchanging conserved microscopic quantities.  At a mesoscopic scale only the energies of the macroscopic bodies are accessible together with  the configurations of the contact system.  We consider a class of models where the contact system, as well as macroscopic bodies, have a finite number of possible configurations. The global system with only discrete degrees of freedom has no
  microscopic Hamiltonian dynamics, but  it is shown that, if the microscopic 
  dynamics   is assumed to be deterministic and ergodic  and to conserve energy according to some specific pattern, and if the mesoscopic evolution of the global system is approximated by a Markov process as closely as possible, then  the
mesoscopic  transition rates obey three constraints. In the limit where macroscopic bodies can be considered as reservoirs at thermodynamic equilibrium (but with different intensive parameters) the mesoscopic transition rates turn into transition rates for the contact system and the third constraint becomes local  detailed balance ; the latter 
   is generically expressed in terms of the microscopic exchange entropy variation, namely the
  opposite of the variation of the thermodynamic entropy of the reservoir involved in
  a given microscopic jump of the contact system configuration.  For a finite-time evolution after contact has been  switched on  we derive a  fluctuation relation for the joint probability of the heat amounts received from the various reservoirs. The generalization to systems exchanging  energy, volume and matter with several reservoirs, with a possible conservative external force acting on the contact system, is given explicitly.

 
 \vskip 0.5cm 
{\it Corresponding author ~:} \\
CORNU Françoise, \\
Fax: 33 1 69 15 82 87 \\
E-mail: Francoise.Cornu@u-psud.fr
\end{abstract}

\clearpage

\section{Introduction}

An archetype of spontaneous irreversible processes that occur when two macroscopic bodies prepared in equilibrium states with different intensive parameters are brought together is thermal contact. The simplest physical situation is when  two bodies ${\cal B}_1$ and ${\cal B}_2$ initially at equilibrium at different temperatures are set into contact through a diathermal incompressible interface which can be either immaterial, the common boundary of the two solid bodies, or a very thin material wall between the two bodies which may be solids or fluids.

First we recall  known results valid under the assumption that the energy of the interactions through the interface is negligible. At a macroscopic level, laws of thermodynamics \cite{Callen1960} predict that in the infinite time limit the  isolated  global system  reaches an equilibrium state where the total thermodynamic entropy   ${\STH_\text{tot}}={\STH_1}+{\STH_2}$ is  maximum  with respect to 
any variation of the internal energies $U_a$  ($a=1,2$) at fixed total internal 
energy\footnote{This  property  can be retrieved  in the framework of equilibrium statistical mechanics, where the thermodynamic entropy  $\STH_\text{tot}$ corresponds to the thermodynamic limit  of the Boltzmann entropy $\SB_\text{tot}$ and the equilibrium fluctuations of the energies in the two bodies are described by the microcanonical probability distribution.} $U_\text{tot}=U_1+U_2$.
 The maximization condition upon ${\STH_\text{tot}}$ entails that  the final inverse thermodynamic temperatures, 
 defined\footnote{In the whole paper  Boltzmann constant $\kB$ is set equal to $1$, the inverse temperature reads $1/T\equiv \beta$
and $S$ denotes a dimensionless entropy. \label{kB}} as $\beta_a\equiv \partial \STH_a / \partial U_a$, are equal, and that    the heat capacity at constant volume of every macroscopic body  is positive. As a consequence the net heat transfer that  leads to equilibrium  flows from the hotter body to the colder one.  
Recently the fluctuations of the heat transfer $\Heat_2$  from body ${\cal B}_2$ towards body ${\cal B}_1$ when  thermal contact is set on during a finite time $t$ have been shown to   obey a fluctuation relation \cite{JarzynskiWojcik2004} : under the assumption  that  only values of  heat amounts $\Heat_2$  that are  far larger than the interaction energy through the interface  are considered,  the probabilities 
to measure either a value $\Heat_2$  or its opposite are linked by 
 $\Prob(\Heat_2;t)=\exp\left[(\beta_1-\beta_2)\Heat_2\right]\Prob(-\Heat_2;t)$. The latter relation implies that $(\beta_1-\beta_2)\Esp{\Heat_{2}}\geq 0$, in agreement with the laws of thermodynamics. 

\vspace{.1cm}

In the present paper we consider the fluctuations of heat exchanges  when  the  interaction energy $\En$ though the interface between the two macroscopic bodies is taken into account. We begin by stating our \textit{basic assumptions}.

$\scriptstyle{I}$) The   contact system $\syst$ has a finite number of configurations $\C$, so that its energies $\En(\C)$ are discretized and bounded.

$\scriptstyle{II}$) The system $\syst$ has no internal dynamics (an expression of its thinness) and its configurations can change only thanks to interactions with the macroscopic bodies ${\cal B}_a$'s  which also have discrete degrees of freedom and which make the  populations in energy levels  of $\syst$ vary\footnote{This is not the case for models that are aimed to study how Fourier law arises when the system that settles thermal contact has a macroscopic width or length  and when the degrees of freedom that are not in contact with the heat sources evolve under a deterministic energy-conserving microscopic dynamics.}. 

$\scriptstyle{III}$) We take it as the definition of the interface that ${\cal B}_a$'s do not interact directly with one another  and we add the physically motivated assumption that no degree of freedom in the interface interacts directly with several ${\cal B}_a$'s. 

$\scriptstyle{IV}$) Discrete degrees of freedom cannot have Hamiltonian evolution but we assume that the dynamics of the global system is ergodic and energy preserving. In a given energy level all ergodic microscopic dynamics have the same period.

We call coarse graining the procedure that goes from a microscopic description of all the degrees of freedom of the global system to a description only of the energies $E_a$ of the macroscopic bodies ${\cal B}_a$ and the configuration $\C$ of the system $\syst$. We call $(\C,E_1,E_2)$ a mesoscopic configuration.  An immediate consequence of $\scriptstyle{IV}$) is that in a coarse grained description the ``probability'' of a mesoscopic configuration  $(\C,E_1,E_2)$ calculated as a time  average over a period of  microscopic dynamics coincides with  the microcanonical probability distribution $\Probmc(\C,E_1,E_2)$. 

\vspace{.2cm}

The main results of the paper are the following.

-- \textit{First}, using our assumptions, in particular $\scriptstyle{IV}$), we show that, during a period of 
microscopic dynamics,  the number of deterministic jumps that occur from one mesoscopic configuration of the global system  to another one  is equal to the number of jumps in the opposite sense. 

-- \textit{Second}, we show that if the mesoscopic dynamics is replaced by the effective Markov process with the same one- and two-body ergodic averages, its transition rates between mesoscopic configurations  $(\C,E_1,E_2)$ satisfy \\
1) \textit{irreducibility} (i.e. the graph associated to the Markov process is connected) \\
2) \textit{microreversibility} (i.e. a transition and its inverse are either both allowed or both forbidden) \\
3) the \textit{microcanonical detailed balance}  with the extra restriction that a single macroscopic body is involved in the transition and its inverse (see \eqref{ratioWmc}).
 \\ We notice that the three properties ensure that  in the infinite-time limit the stochastic evolution  does lead to a unique stationary state which coincides with  the microcanonical probability distribution $\Probmc(\C,E_1,E_2)$.

 As larger and larger macroscopic bodies are considered, one can identify longer and longer time windows  during which they can be considered as being in an equilibrium state, i.e. their thermodynamic entropies are well defined and their temperatures  are constant (but different in general) over the whole time window. In this  regime, a Markov dynamics for the system $\syst$ emerges,  the irreducibility and microreversiblity conditions are preserved   and the mesoscopic microcanonical detailed balance for transitions among $(\C,E_1,E_2)$'s  turns into the so-called ``local '' detailed balance\footnote{Nowadays the denomination ``local'' detailed balance seems to be the standard one used  in the literature for discrete state systems driven out of equilibrium under various kinds of external constant conditions (see below). In the present thermal contact setting, the term ``local'' may be viewed as referring to the local interaction between the contact system and each energy reservoir, but the  denomination ``generalized'' detailed balance is also to be found in Ref.\cite{Derrida2007}.}
 for transitions among $\C$'s (see  \eqref{ergodicThermo}). 
 The latter has been often used in the literature as a basic assumption in the description of the stochastic evolution of a system under energy  exchanges with several  thermostats at different temperatures (see comments below).

-- \textit{Third}, we use the local  detailed balance to exhibit an exact finite time fluctuation relation for the joint probability distribution of the heat amounts received by the interface system $\syst$ from both macroscopic bodies  in a protocol where thermal contact is set on at the initial time (see text before \eqref {ratioProbQ1Q20}).

-- \textit{Fourth}, we show how the above considerations generalize to cases when not only energy but also particles and/or volume  are exchanged among macroscopic bodies via an interface. The number of states of the interface (and in particular, the energies and/or particle contents and/or volumes it can have) are still assumed to be finite. The case where a conservative external force acts on some global coordinate of the contact system is included in the description.

\vspace{.2cm}

A slightly extended description will allow us to situate our work in the already available litterature.

The issue of how to get a valuable model of thermal contact can also be addressed not by assuming some 
properties for  the microscopic dynamics but by  imposing some constraints directly upon the mesoscopic transition rates in order to ensure that  the stochastic evolution of the global system does lead to equilibrium in the infinite-time limit\footnote{The latter point of view 
 is that used by Glauber  in his investigation of the time-dependent statistics of the Ising spin chain when spin flips are supposed to be generated by energy exchanges with a thermostat at the inverse temperature $\beta$ :   he imposes that   in the infinite-time limit  the corresponding stochastic dynamics    leads to the canonical equilibrium probability at inverse temperature $\beta$ with the Ising Hamiltonian \cite{Glauber1963}.}.
 The uniqueness of the stationary state is guaranteed if the transition rates obey the irreducibility and microreversibility conditions, and it coincides with a prescribed equilibrium state if the transition rates obey the detailed balance with the corresponding equilibrium probability distribution. In the case of thermal contact the latter  distribution is the microcanonical probability $\Probmc(\C,E_1,E_2)$  and this approach is to be found in Ref.\cite{Derrida2007}. However,  we stress that, because of the interaction pattern, the microcanonical detailed balance \eqref{ratioWmc}  arising in our derivation from the properties of the microscopic dynamics involves in fact only a single macroscopic body at a time, as explicitly written in \eqref{ergodic}.

As for the local  detailed balance  \eqref{ergodicThermo} for thermal contact, valid in the  regime where the macroscopic bodies remain at equilibrium, some derivations from the existence of an underlying Hamiltonian microscopic dynamics   invariant under time reversal   can be found in the literature. In particular there are some similarities between our microscopic approach and the derivation by Maes and Netocny \cite{MaesNetocny2003}, who use (albeit in a different order, in the context of Hamiltonian dynamics and using time reversal) coarse graining, and, after postulating that the macroscopic bodies are in  steady states, a Markovian approximation. They also  use a prescription analogous to the interaction pattern for our jump dynamics, since the reservoirs are assumed to be spatially separated, each being in contact only with the system.  This spatial separation and local coupling to the system ensures that the the variations of the reservoir variables are  determined only  by the the variations of the system variables \cite{Maes2003Poincare}.
In fact this fundamental property of the contact setting has also been used in an earlier derivation of the local  detailed balance in Ref.\cite{BergmannLebowitz1955}\footnote{In Ref.\cite{BergmannLebowitz1955}  the  transition rates appear 
in a stochastic integral  term that is added to the  Liouville equation for  the Hamiltonian evolution of the probability distribution of the small system  in its phase space. The stochastic term is intended to mimic the effect of the instantaneous  elastic collisions between the small system and the degrees of freedom of reservoirs that have a  canonical equilibrium probability distribution. The  argument relies on the assumption of the absence of  correlation before a collision and  on the time reversal invariance of both the collision process and  the Hamiltonian evolution of the system when it is isolated.}.

As for the generalization of the microscopic derivation of the local  detailed balance to processes where the underlying microscopic dynamics allows  exchanges of various microscopically conserved quantities (energy and/or matter and/or volume), the key observation is the following. Once  the mesoscopic microcanonical  detailed balance \eqref{ratioWmcgen}, with the constraint that only one macroscopic body is involved in a given configuration jump of the interface $\syst$, has been derived  then, as already pointed out in Ref.\cite{Derrida2007},  its replacement by the  local  detailed balance in the regime where macroscopic bodies behave as  reservoirs in different thermodynamical equilibrium states corresponds to the replacement of  the variation of the Boltzmann entropy  of the macroscopic body involved in the configuration jump by the  variation of  its thermodynamic entropy at fixed intensive parameters. 

Therefore the local  detailed balance  can be expressed in a generic way (see \eqref{MDBexch})  in terms of the exchange entropy variation\footnote{\label{exchangename} Exchange entropy variation is an abbreviation for ``variation of entropy due to  exchanges'' of energy  with a thermal bath, or more generally, of some measurable conserved quantities with various reservoirs.} 
$\deltaexch S(\C'\leftarrow\, \C)$ associated with a jump  of the small system from a microscopic configuration  $\C$ to another one $\C'$, which is defined  as the opposite of the infinitesimal variation of the   thermodynamic entropy of the reservoir ${\cal B}_a$ that  causes the jump of configuration from $\C$ to $\C'$ by  exchanging energy and/or volume and/or matter with the system $\syst$. Its explicit expression is given in \eqref{ExplicitDeltaSexchGen} with some comments about its physical content. It may involve a conservative force acting on the system.
In fact the local  detailed balance has been used in various specific forms :  for instance for the description of exchanges  of  particles (energy quanta)  with two particle (energy) reservoirs at different chemical potentials (temperatures) at the boundaries of a one-dimensional lattice where particles move stochastically according to the rules of the symmetric simple exclusion process \cite{BodineauDerrida2007}, or for coupled exchanges of energy and particles in molecular motor models (see among others \cite{AndrieuxGaspard2006PRE,SchmiedlSpeckSeifert2007,LiepeltLipowsky2007}). 
The present derivation allows to describe a mobile thin diathermal wall (with a possible conservative force acting on it) separating a vessel in two parts filled with gases at different temperatures and pressures.

We notice that for an another class of non-equilibrium processes  driven by constant external constraints, namely particle currents driven by a constant non-conservative force acting on  particles in contact with a single heat bath (for instance interacting particles moving stochastically on a ring in some uniform  field) the issue of a microscopic derivation of the corresponding local detailed balance assumed in Ref.\cite{KatzLebowitzSpohn1984} has been addressed in 
Ref.\cite{Tasaki2007}. Then the local detailed balance has the same form  as a canonical detailed balance, but the stationary state is not the   canonical equilibrium state because of the periodic boundary conditions.

We consider the  following protocol for  thermal contact  :
 the system $\syst$ is a  thin piece of material prepared at inverse temperature $\beta_0$  and  at the initial time of measurements it is inserted between two  macroscopic bodies at equilibrium at different inverse temperatures $\beta_1$ and $\beta_2$  (and which remain in their equilibrium state during the whole  protocol). Then the transition rates of the stochastic evolution of $\syst$ satisfy the  local  detailed balance and  the joint probability of the   heats $\Heat_1$ and $\Heat_2$ received by $\syst$ from both macroscopic bodies  during a time $t$ after contact has been set on is shown to obey  relation  \eqref{ratioProbQ1Q2}. The  exchange entropy variation associated with the  heats $\Heat_1$ and $\Heat_2$ received by $\syst$ from both thermostats reads $\Deltaexch S(\Heat_1,\Heat_2)= \beta_1 \Heat_1+\beta_2 \Heat_2$, and  the detailed fluctuation relation  can be written   more explicitly as
\be
\label{ratioProbQ1Q20}
\frac{\Prob\left(\Heat_1,\Heat_2; t\right)}{\Prob
\left(-\Heat_1,-\Heat_2; t\right)}=e^{(\beta_0-\beta_1)\Heat_1+(\beta_0-\beta_2)\Heat_2}.
\ee
It has been checked in a solvable model for thermal contact \cite{CornuBauerBETA}.
Moreover the corresponding integral fluctuation relation,  
 $\Esp{\exp\left[(\beta_1-\beta_0)\Heat_1+ (\beta_2-\beta_0)\Heat_2\right]}$ $=1$, where $\Esp{\ldots}$ denotes an expectation value when the experiment is repeated a large number of times, 
entails through Jensen's inequality that the mean heat amounts that are exchanged  from the initial time where the thermal contact is set on  must satisfy 
\be
\label{ClausiusLawRetrieved} 
(\beta_0-\beta_1) \Esp{\Heat_1(t)}+(\beta_0-\beta_2) \Esp{\Heat_2(t)}\geq 0.
\ee
In fact, by conservation of energy, $\Esp{\Heat_1(t)}+\Esp{\Heat_2(t)}$ is equal to the difference between the mean energies of the system $\Esp{\En}_t-\Esp{\En}_{t=0}$, and the inequality \eqref{ClausiusLawRetrieved}  also reads 
$(\beta_1-\beta_2) \Esp{\Heat_2(t)}\geq (\beta_1-\beta_0) \left[\Esp{\En}_t-\Esp{\En}_{t=0}\right]$. 
We notice that the fluctuation relation \eqref{ratioProbQ1Q20} is compatible with  the relation quoted at the beginning of the introduction and derived in Ref.\cite{JarzynskiWojcik2004} for a similar protocol : in the latter  no material contact system  $\syst$ is inserted between the macroscopic bodies, but  the global system is assumed to have an underlying microscopic Hamiltonian dynamics (invariant under time reversal)  whose Hamiltonian contains an interaction term only when the two macroscopic bodies are in contact. (Remarks in subsection \ref{Ergodicity} sustains the analogy between both protocols). The initial equilibrium states of the macroscopic bodies are described in the canonical ensemble, no assumption is made about the evolution of the temperatures of the macroscopic bodies, but 
 only heat amounts  that are  far larger than the variation of the interaction energy through the interface between the  initial and final  times of contact  are considered, which  in our setting is equivalent to consider only values  $\vert\Heat_2 \vert \gg \vert\Esp{\En}_t-\Esp{\En}_{t=0}\vert$.

For the larger class of models where reservoirs exchange microscopically conserved quantities and for which the microscopic derivation of the local  detailed balance can  be performed,
 the exchange entropy variation along a history of the system, $\Deltaexch S$, is the crucial quantity at the root
 the fluctuation relations \eqref{DFRJointProbChi} and  \eqref{DFRDeltaexchexcess0gen}, which are generalizations of the pure thermal contact relations  \eqref{ratioProbQ1Q2} and  \eqref{DFRDeltaexchexcess} respectively.
 The protocol for setting contact is such that the latter relations, as well as the  exchange entropy variation $\Deltaexch S$, are expressed only in terms of  the  quantities that are exchanged between the interface and every  reservoir  and on the intensive equilibrium parameters of the reservoirs. All these quantities are  experimentally measurable quantities. On the contrary the  probability distribution of the system configurations is very hard to determine from experiments. We recall that the probability distribution of the system configurations is involved in  the variation of the system Shannon-Gibbs entropy which appeared in the seminal works by Crooks \cite{Crooks1998, Crooks1999, Crooks2000} and has led to the notion of entropy production along a stochastic trajectory of the system microscopic configurations  (see \cite{Seifert2005,Seifert2008}). 

Finally we point out that the action functional introduced by Lebowitz and Spohn \cite{LebowitzSpohn1999} in terms of the transition rates of a Markov process satisfying only the irreducibility constraint \eqref{Mirreducible0} and the microreversiblity constraints \eqref{MicroRevCond0} coincides with the opposite of the exchange entropy variation when the local  detailed balance condition \eqref{MDBexch} is also met. As a consequence, fluctuations relations in the long-time limit (which are out of the scope of the present paper) for $\Deltaexch S$ and for the cumulative exchange quantities can be derived respectively from the generic results in Refs.\cite{LebowitzSpohn1999} and \cite{AndrieuxGaspard2007JStatPhys} or  retrieved from the present finite time fluctuation relations.
A recent
extended review about the larger frame of stochastic thermodynamics is to be found in Ref.\cite{Seifert2012}, and its formulation in  the specific class of Markov jump processes  in continuous time is reviewed  in Ref.\cite{HarrisSchutz2007}.

\vspace{.2cm}

The paper is organized as follows. In section \ref{sectionConstraints}  constraints upon the transition rates are derived from the assumptions about the microscopic dynamics of the whole system. Our first result is derived  in  subsection \ref{Ergodicity} and Appendix \ref{CoarseGrainedProperty}. The prescription to approximate the  mesoscopic dynamics  by a Markov process is described in Appendix  \ref{Markovapprox} and our second result about the microcanonical detailed balance is derived in  subsection \ref{MarkovianApproximationConseq}. Our argument involves no time reversal symmetry, and, for a comparison,  the  microcanonical detailed balance  is rederived in the case of an underlying Hamiltonian dynamics invariant under time reversal in appendix \ref{TimeReversal}. Its transformation into the local  detailed balance in the limit where the sizes of the macroscopic bodies go to infinity before the time evolution of the interface system is considered  is described in subsection \ref{TransientSec}.
 In section \ref{FiniteTimeSymmetry} we exhibit the  derivation of the fluctuation relation \eqref{ratioProbQ1Q20} and that of the corresponding fluctuation relation for the ``excess'' exchange entropy variation. In section \ref{GeneraliztionSeveralCurrents} the key ingredients used in the previous two sections are generalized to the case of exchanges involving also variations of volume or number of particles.

\section{Constraints upon transitions rates}

\label{sectionConstraints}

In the present section we review some of the constraints that ergodic
deterministic energy-conserving microscopic dynamics puts on the statistical
mesoscopic description of a finite system $\syst$ which establishes thermal contact
between energy reservoirs ${\cal B}_a$'s with $a=1,\ldots,A$.

Indeed, the following situation occurs commonly : the interactions in the whole
system allow to define one small part $\syst$ in contact with otherwise independent
large parts ${\cal B}_a$'s.  The large parts, which involve a huge number of degrees of
freedom, do not interact directly among each other (this gives a criterion to
identify the distinct large parts), but are in contact with the small part,
which involves only a few degrees of freedom.  Moreover each degree of freedom
in the small part is directly in contact with at most one large part and can vary  only through its interaction 
with the latter large part. This
results in a star-shaped interaction pattern.  It is convenient then to forget
about the microscopic description of the large parts, and turn to a statistical
description of their interactions with the small part. Some general features of
the statistical description can be inferred from microscopic ergodicity.

\subsection{Ergodicity}
\label{Ergodicity}

\subsubsection{Ergodicity in  classical Hamiltonian dynamics}
\label{HamiltonianErgodicity}

In classical mechanics, the time evolution of a system in phase space is
described by a Hamiltonian $H$. If the system is made of several interacting
parts, the Hamiltonian is then referred to as the total Hamiltonian, $H\equiv H_\text{tot}$, and it splits as $H_\text{tot}=H_\text{dec} +H_\text{int} $, where
$H_\text{dec} $ accounts for the dynamics if the different parts were decoupled
and $H_\text{int} $ accounts for interactions. The energy hypersurface $H_\text{tot}=E$,
usually a compact set, is invariant under the time evolution, and in a generic
situation, this will be the only conserved quantity.

The ergodic hypothesis
states that a generic trajectory of the system will asymptotically cover the
energy hypersurface uniformly. To be more precise, phase space is endowed with
the Liouville measure (i.e. in most standard cases the Lebesgue measure for the
product of couples made by every coordinate and its conjugate momentum), which
induces a natural measure on the energy hypersurface, and ergodicity means that,
in the long run, the time spent by the system in each open set of the energy
hypersurface will be proportional to its Liouville measure. Ergodicity can
sometimes be built in the dynamics, or proved, but this usually requires immense
efforts.

Ergodicity depends crucially on the fact that the different parts are coupled :
if $H_\text{int} =0$, each part will have its energy conserved, and motion will
take place on a lower-dimensional surface. If $H_\text{int} $ is very small, the
system will spend a long time very nearby this lower-dimensional surface, but, at
even longer time scales, ergodicity can be restored. By taking limits in a
suitable order (first infinite time and then vanishing coupling among the parts)
one can argue that the consequences of ergodicity can be exploited by reasoning
only on $H_\text{dec}$.

Notice that in this procedure, we have in fact some kind of dichotomy : $H_\text{dec}
$ defines the energy hypersurface, but cannot be used to define the ergodic
motion, which is obtained from $H_\text{tot}=H_\text{dec} +H_\text{int} $ via a limiting
procedure.  So the dynamics conserves $H_\text{dec} $ but is not determined by
$H_\text{dec} $.

\subsubsection{Ergodicity in  deterministic  dynamics for discrete variables}
\label{ErgodicityDiscrete}

Our aim is to translate the above considerations in the context of a large but
finite system described by discrete variables such as classical Ising spins.

In the case of discrete dynamical variables, one can still talk about the energy
$E_\text{tot}$ of a configuration, but there is no phase space and no Hamiltonian dynamics
available.  So there is no obvious canonical time evolution. This is where we
exploit the previously mentioned dichotomy: we do not define the time evolution
in terms of $E_\text{tot}$, but simply impose that the deterministic time evolution
preserves $E_\text{dec}$ and that it respects the  star-shaped interaction pattern between the
small part and the large parts. Besides we also impose that the time evolution is ergodic.

We consider that time is discrete as well, because  ergodicity is most simply expressed in discrete time. Then deterministic dynamics is given by a bijective
map, denoted by $\Udisc$ in what follows, on configuration space, applied at
each time step to get a new configuration from the previous one. 
As the
configuration space is finite, the trajectories are bound to be closed. Then, for a given initial value $E$ of $E_\text{dec}$, a specific dynamics $\Udisc$ conserving $E_\text{dec}$ corresponds to a periodic evolution of the microscopic configuration of the full  system inside the energy level $E_\text{dec}=E$.

Ergodicity entails that the corresponding closed trajectory covers fully the energy level 
$E_\text{dec}=E$, and then it must cover it exactly once during a period because the dynamics is
one to one. As a consequence the period of the ergodic evolution inside a given energy level 
of $E_\text{dec}$ is the same for all choices of ergodic dynamics $\Udisc$ that conserves $E_\text{dec}$. This period, denoted by $N$ in time step units, depends only on the value $E$ of $E_\text{dec}$,
\be
\label{valuePeriod}
N=\Omega_\text{dec}(E),
\ee
where  $\Omega_\text{dec}(E)$ is the total number of microscopic configurations in the level $E_\text{dec}=E$.
This is reminiscent of the microcanonical ensemble. Let us note that
in classical mechanics, there is a time reversal symmetry, related to an
involution of phase space, changing the momenta to their opposites while leaving
the positions fixed. In the discrete setting, involutions $J$ such that
$J\Udisc J=\Udisc ^{-1}$ always exist, but there is no obvious candidate among
them for representing time reversal and allowing to draw conclusions from it.

In the  context of discrete variables  the star-shaped interaction pattern is implemented as follows. We may naively assume that the energy
conserved by the dynamics $\Udisc$ is simply $E_\text{dec}$, as if there were no energy
for the interactions between the small part and the large ones, but  $\Udisc$ must
reflect the fact that the large parts interact only indirectly: there is an
internal interaction energy $\En(\C)$ for every configuration $\C$ of the small
part and each change in the small part can be associated with an elementary
energy exchange with one of the large parts. If the small part can jump from
configuration $\C$ to configuration $\C'$ in a single time step by exchanging
energy with large part ${\cal B}_a$, we use the notation $\C'\in\F_a(\C)$. In this
configuration jump the energy $E_\text{dec}$ of the global system is conserved and the energy
of  large part ${\cal B}_a$ is changed from $E_a$ to $E'_a$ according to the
conservation law
\be
 \label{conserE}
E'_a-E_a=
\begin{cases}
  -\left[\En(\C')-\En(\C)\right]   &\textrm{if}\quad  \C'\in\F_a(\C) \\
  0 &\textrm{otherwise},
\end{cases}
\ee 
while the energies of the other large parts are unchanged. Apart from these energy exchange constraints and from ergodicity, the deterministic dynamics $\Udisc$ is  supposed to obey some other natural physical constraints which will be specified later (see subsection \ref{MarkovianApproximationConseq}).

As a final remark, we mention how, in a very simple case,  some kind of deterministic map $\Udisc$ that preserves $E_\text{dec}$ and obeys the star-shaped interaction pattern can be associated with  a deterministic map  $\widetilde{\Udisc}$ that conserves $E_\text{tot}$. We consider the case where the energy exchange between every large part ${\cal B}_a$ and the small part is ensured by an interaction energy $E_\text{int}^{(a)}$ between a classical spin 
$\sigma_a^\star$ in part ${\cal B}_a$ and a classical spin $\sigma_a$ in the small part and we consider only maps $\widetilde{\Udisc}$ that not only conserve  $E_\text{tot}$ but also satisfy the following rules for all large parts :  (1)  spins $\sigma_a^\star$ and $\sigma_a$ are always flipped at successive time steps (in an order depending on the precise dynamics $\widetilde{\Udisc}$) ; (2) the variations of the interaction energy $E_\text{int}^{(a)}$ associated  with these successive two flips are opposite to each other. Then the map $\Udisc$ that conserves $E_\text{dec}$ is defined from the map $\widetilde{\Udisc}$ by merging  every pair of time steps where $\sigma_a^\star$ and $\sigma_a$ are successively flipped into a single time step where $\sigma_a^\star$ and $\sigma_a$ are  simultaneously flipped. Indeed, in the latter pair of time steps of $\widetilde{\Udisc}$, by virtue of hypothesis (2),  the successive two variations of $E_\text{int}^{(a)}$ cancel each other and the variation of  $E_\text{tot}$ after these two time steps coincides with the variation of $E_\text{dec}$, since by definition the latter variation is $\Delta E_\text{dec}=\Delta E_\text{tot}- \Delta E_\text{int}^{(a)}$. 
Therefore, if map $\widetilde{\Udisc}$ conserves $E_\text{tot}$ at every time step, then the corresponding map $\Udisc$ where the latter  successive two flips occur in a single time step preserves   $E_\text{dec}$ : the conservation rule \eqref{conserE} is indeed satisfied. The suppression of time steps in the procedure that defines $\Udisc$ from  $\widetilde{\Udisc}$ corresponds to a modification of the accessible configuration space that reflects the fact that  the energy level $E_\text{tot}=E$ and $E_\text{dec}=E$ do not coincide.

\subsubsection{Constraints from ergodicity and interaction pattern upon coarse-grained evolution}

We start from the the familiar observation that keeping
track of what happens in detail in the large parts is out of our abilities, and
often not very interesting anyway. Our ultimate interest is in the evolution of the configuration $\C$ of
the small part in an appropriate limit. As an intermediate step, we keep track
also of the energies $E_a$'s  in the large parts, but not of the detailed configurations
in the large parts.

The coarse graining that  keeps track only of the time evolution of the configuration $\C$ of the small part and the energies $E_a$'s of the large parts is defined as follows.
With each microscopic configuration $\xi$  of the full system we
can associate the corresponding configuration $\C=\C(\xi)$ of the small part and
the corresponding energy $E_a=E_a(\xi)$ carried by part ${\cal B}_a$. To simplify the notation, we let $\underline{E}$ denote
the collection of $E_a$'s, so $\underline{E}$ is a vector with as many
coordinates as there are large parts. 

As shown in previous subsubsection,  if  the  initial value of the energy $E_\text{dec}$ is equal  to $E$, then, over a period equal to $N=\Omega_\text{dec}(E)$ in time step units, the trajectory of $\xi$ under any  microscopic ergodic dynamics $\Udisc$ corresponds to a cyclic permutation of all the microscopic configurations  in the  energy level $E_\text{dec}=E$. 
Thus ergodicity entails that at the coarse grained level,  if $N_{(\C,\underline{E})}$ denotes the occurrence number  of
$(\C,\underline{E})$ during the  period of $N$ time steps,
$N_{(\C,\underline{E})}$ is nothing but 
$\Omega_\text{dec}(\C, \underline{E})$, the number
of microscopic configurations of the full system when the small part is in configuration $\C$ and the large parts have energies $E_a$'s in the energy level $E=E_\text{dec}$, namely
\be
\label{Nxvalue}
N(\C, \underline{E})=\Omega_\text{dec}(\C, \underline{E})
\ee
with
\be
\label{EnergyConstraint}
E=E_\text{dec}\equiv\En(\C)+\sum_a E_a.
\ee
In the following we fix the value  of $E$ and  $N$ is called the period of the dynamics, while the  energy constraint  $E=\En(\C)+\sum_a E_a.$ is often implicit in the notations.

Another crucial point is that, since any specific dynamics $\Udisc$ under consideration respects both the conservation of $E_\text{dec}$ and the interaction pattern specified at the end of subsubsection \ref{ErgodicityDiscrete},   the number
of jumps from $(\C,\underline{E})$ to $(\C',\underline{E'})$ over the  period of $N$ time steps, denoted by 
$N_{(\C,\underline{E}),(\C',\underline{E}')}$, is equal to the number of the reversed jumps 
from $(\C',\underline{E'})$ to $(\C,\underline{E})$ during the same time interval, namely
\be
\label{NumbersEqual}
N_{(\C,\underline{E}),(\C',\underline{E}')}=
N_{(\C',\underline{E}'),(\C,\underline{E})}. 
\ee 
In Appendix \ref{CoarseGrainedProperty}
we  give graph-theoretic conditions, not related to ergodicity, that ensure
this property, and show that they are fulfilled in one relevant example, as a consequence of the star-shaped interaction pattern.

\subsection{Markovian approximation for the  mesoscopic dynamics}
\label{MarkovianApproximationConseq}

\subsubsection{Definition of a Markovian approximation for the  mesoscopic dynamics}
\label{DefinitionMarkovApprox}

For our purpose, we first rephrase the coarse-grained evolution as follows. As already noticed, over the period of $N=\Omega_\text{dec}(E)$ time steps, a trajectory under any  microscopic ergodic dynamics $\Udisc$ in the energy level $E_\text{dec}=E$ corresponds to a cyclic permutation of  the N microscopic configurations $\xi$'s in the  energy level. Therefore, if the configuration at some initial time is denoted by  $\xi_1$, then the trajectory  is represented by the sequence
$\omega=\xi_1\xi_2\cdots \xi_N$ where $\xi_{i+1}=\Udisc \xi_i$ with
$\xi_{N+1}= \xi_1$. By the coarse-graining procedure that retains only the mesoscopic variable  $x\equiv(\C,\underline{E})$, the succession of distinct microscopic configurations $\omega$ is  replaced by  $w=x_1x_2\cdots x_N$ where
$x_i=x(\xi_i)=(\C(\xi_i),\underline{E}(\xi_i))$. 
In  $w$ various $x_i$'s take the same value, and a so-called transition corresponds to the case $x_{i+1}\not=x_i$,  namely the case where the configuration $\C$ of the small system is changed in the jump of the microscopic configuration of the full system from $\xi_i$ to  $\xi_{i+1}=\Udisc \xi_i$.

Since the large parts involve many degrees of freedom, the number of times some given value  $x=(\C,\underline{E})$ appears  in the coarse-grained sequence  $w$ is huge, and
even if $\omega$ is given by the deterministic rule $\xi_{i+1}=\Udisc \xi_i$,
there is no such rule to describe the sequence $w$. Moreover the  microscopic configuration 
at the initial time, $\xi_1$, is not known  so that  the coarse-grained sequence that actually appears in the course of time is in in fact a  sequence deduced from $w$ by a translation of all indices.

As explained in Appendix \ref{DiscreteTMarkov}, one may associate to the sequence $w$
a (discrete time) Markov chain such that the mean occurrence frequencies  of the patterns $x$ and $xx'$ in a stationary stochastic sample are equal to the corresponding values, $N_x/N$ and  $N_{xx'}/N$, in the sequence $w$ determined by the dynamics $\Udisc$ (up to a translation of all indices corresponding to a different value of the initial microscopic configuration). Whether this Markovian effective description is accurate depends on several things: the choice of $\Udisc$, the kind of statistical properties of $w$ one wants to check, etc.

We may also argue (see Appendix \ref{ContinousTimeApproximation}) that a continuous time description is enough if we restrict our attention to microscopic dynamical maps $\Udisc$'s such  that transitions, namely the  patterns $xx'$ with $x'\not= x$, are rare and of comparable mean occurrence frequencies over the  period of $N$ time steps. In other words, most of the steps in the dynamics amount to reshuffle the
 configurations of the large parts without changing their energies, leaving the
 configuration of the small part untouched. The latter physical constraint and the hypothesis of  the validity of the Markovian approximation  select a particular class of dynamics $\Udisc$.

With these assumptions,  we associate to the sequence $w$ of coarse grained variables 
$(\C, \underline{E})$ a Markov process  whose stationary measure shares some  of the
 statistical properties  of $w$, namely the values of the mean occurrence frequencies of length $1$ and length $2$ patterns. The  transition rate from  $(\C,\underline{E})$ to 
 $(\C',\underline{E}')$ with $(\C,\underline{E}) \neq (\C',\underline{E}')$ in the approximated Markov process  is given by \eqref{Wxxprim}, where we just have to make the substitutions $N_x=N_{(\C,\underline{E})}$ and 
 $N_{xx'}=N_{(\C,\underline{E}),(\C',\underline{E}')}$, with the result
 \be
\label{TauxMesoW0}
W(\C',\underline{E}'\leftarrow \C,\underline{E})=
\frac{N_{(\C,\underline{E}),(\C',\underline{E}')}}{\tau \,N_{(\C,\underline{E})}}\quad
\text{ for } (\C,\underline{E}) \neq (\C',\underline{E}'),
\ee 
where $\tau$ is a time scale such that  $W(\C',\underline{E}'\leftarrow \C,\underline{E})$ is of order unity. 
The corresponding stationary distribution  is given by \eqref{propequiv1},  
\be
\label{Probmicro0}
\Probst^W(\C,\underline{E})=\frac{N_{(\C,\underline{E})}}{N}.
\ee 
We recall that  $N=\sum_{(\C,\underline{E})}N_{(\C,\underline{E})}$ and 
$N_{(\C,\underline{E})}=\sum_{(\C',\underline{E'})}N_{(\C,\underline{E}),(\C',\underline{E}')}$.

\subsubsection{Microcanonical detailed balance and other properties}
\label{Microcansubsubsection}

By virtue of the ergodicity property 
 \eqref{Nxvalue} at the coarse-grained level,  the transition rate in the approximated Markov process reads
 \be
\label{TauxMesoW}
W(\C',\underline{E}'\leftarrow \C,\underline{E})=
\frac{N_{(\C,\underline{E}),(\C',\underline{E}')}}{\tau \,\Omega_\text{dec}(\C, \underline{E})}\quad
\text{ for } (\C,\underline{E}) \neq (\C',\underline{E}').
\ee
 Meanwhile, by virtue of the ergodicity properties  \eqref{valuePeriod}  and \eqref{Nxvalue}, the corresponding stationary distribution 
 \eqref{Probmicro0} is merely the microcanonical distribution
\be
\label{Probmicro}
\Probst^W(\C,\underline{E})=\frac{\Omega_\text{dec}(\C, \underline{E})}{\Omega_\text{dec}(E)}\equiv\Probmc(\C, \underline{E}).
\ee

Ergodicity also entails that all microscopic configurations $\xi$'s in the energy level appear in the sequence $\omega$, and therefore  all possible values of $x$ also appear in the coarse-grained sequence $w$ : so  any mesoscopic state $(\C,\underline{E})$ can be reached from any other mesoscopic state $(\C',\underline{E}')$ by a succession of elementary transitions, even if they are not involved in an elementary transition (i.e. if $N_{(\C,\underline{E}),(\C',\underline{E}')}=0$);
in other words the graph associated with the transition rates $W(\C',\underline{E}'\leftarrow \C,\underline{E})$ is connected, or,
equivalently, the Markov matrix defined from the transition rates is \textit{irreducible}.

The constraint \eqref{NumbersEqual} imposed by the interaction pattern upon the coarse-grained evolution over a period of $N$ time steps entails that the transition rates of the approximated  Markov process given by \eqref{TauxMesoW} obey two properties. First,
 \be
\label{ergodicRev}
W( \C',\underline{E}' \leftarrow \C,\underline{E})\quad \text{and}\quad
W( \C,\underline{E} \leftarrow \C',\underline{E}') 
\ee 
are
either both $=0$ or both $\neq 0$. This property may be called \textit{microreversibility}.
Second, if the transition rates do not vanish, they obey the equality
\be
\label{ratioWmc}
 \frac{W(\C',\underline{E}'\leftarrow \C,\underline{E})}
  {W(\C,\underline{E}\leftarrow \C',\underline{E}')}=
\frac{\Omega_\text{dec}(\C',\underline{E}')}
{\Omega_\text{dec}(\C,\underline{E})}
=\frac{\Probmc(\C',\underline{E'})}{\Probmc(\C,\underline{E})}.
\ee
Observe that the arbitrary time scale $\tau$ has disappeared in this equation.
The equality between the ratio  of transition rates and the ratio of  probabilities in the corresponding stationary distribution is the so-called detailed balance relation. Here the 
stationary distribution is that of the microcanonical ensemble, and we will refer to relation 
\eqref{ratioWmc} as the \textit{microcanonical detailed balance}.

In appendix \ref{TimeReversal}  we rederive a  microcanonical detailed balance similar to \eqref{ratioWmc}  when the underlying microscopic  dynamics is Hamiltonian and invariant under time reversal. (Similar arguments can be found in derivations which rely on different assumptions in Refs.\cite{Wigner1954,vanKampen1992}.) The evolution of the probability distribution of the mesoscopic  variables is approximated by a Markov process according to the same scheme as that introduced in subsubsection \ref {DefinitionMarkovApprox}.
Eventually the comparison between the ways in which the microcanonical detailed balance arises in that case and in our previous argument  is the following.

- When the microscopic variables are continuous coordinates in phase space and evolve according to a Hamiltonian dynamics, in the framework of statistical ensemble theory  the stationary measure for the  mesoscopic variables  is the measure that is preserved under the microscopic dynamics ; the fact that it coincides with the microcanonical distribution is enforced by   the invariance of the Liouville measure in phase space under the Hamiltonian evolution ; the microcanonical detailed balance   for mesoscopic variables that are even functions of microscopic momenta mainly arises from the  invariance under time reversal of the  trajectories in  phase space  (see \eqref{ProbJointHBis}).

- When the microscopic dynamical variables are discrete and evolve under an  energy-conserving map,  the stationary measure for  mesoscopic variables is  defined as the average over a period of the microscopic dynamics ; the fact that it is equal to the microcanonical distribution arises from the ergodicity imposed on the microscopic map $\Udisc$ (see \eqref{valuePeriod} and  \eqref{Nxvalue}) ; the microcanonical detailed balance emerges from  the  equality between the  frequencies of a given transition and the reversed one (over the period needed for  the microscopic  map to cover the energy level),  equality which is enforced by the  star-shaped interaction pattern (see \eqref{NumbersEqual}).

\subsubsection{Further consequence of the interaction pattern}

 In the interaction pattern large parts do not interact directly with one another and $\Omega_\text{dec}(\C, \underline{E})=\prod_a \Omega_a(E_a)$, where $\Omega_a(E_a)$ denotes the number of configurations in large part ${\cal B}_a$ with energy
$E_a$ when it is isolated. Moreover the energy of a single large part is changed  in a given transition, so  if $(\C',\underline{E}')$ is obtained from
$(\C,\underline{E})$ by an energy exchange with part ${\cal B}_a$ that makes $\C$ jump to 
$\C'$ we have the result, with the notation introduced in \eqref{conserE},
\be
\label{SimplifiedRatio}
\textrm{if $\C'\in\F_a(\C)$} \quad
\frac{\Omega_\text{dec}(\C',\underline{E}')}
{\Omega_\text{dec}(\C,\underline{E})}= \frac{\prod_b
  \Omega_b(E'_b)}{\prod_b \Omega_b(E_b)}=\frac{\Omega_a(E'_a)}{\Omega_a(E_a)}.
\ee 
We have used the energy conservation rule  $E'_b =E_b -\delta_{a,b}\left[\En(\C')-\En(\C)\right]$, so
that for $b \neq a$ the multiplicity factors are unchanged in the transition.

The latter ratio of microstate numbers  can be expressed in terms of
the Boltzmann entropies when each part ${\cal B}_a$ is isolated. The dimensionless
Boltzmann entropy (see footnote \ref{kB})
$\SB_a(E_a)$ for the isolated part ${\cal B}_a$ when its energy is equal to $E_a$ is
defined by $\Omega_a(E_a)\equiv\exp \SB_a(E_a)$. With these notations, 
if the transition rate $W( \C',\underline{E}' \overset{\F_a}{\leftarrow}
  \C,\underline{E})$, where $\C'\in\F_a(\C)$, is nonzero, then the transition rate for the reversed jump $W( \C,\underline{E} \overset{\F_a}{\leftarrow} \C',\underline{E}')$, where $\C\in\F_a(\C')$, is also non zero (see \eqref{ergodicRev}) and, by virtue of \eqref{SimplifiedRatio} the relation \eqref{ratioWmc} is reduced to
\be
\label{ergodic}
\frac{W( \C',\underline{E}' \overset{\F_a}{\leftarrow}
  \C,\underline{E})}{W( \C,\underline{E} \overset{\F_a}{\leftarrow}
  \C',\underline{E}')}=\frac{\Omega_a(E'_a)}{\Omega_a(E_a)}
 \equiv e^{\SB_a(E'_a)-\SB_a(E_a)}.
\ee
The latter formula is the first important stage of the argument.

We stress that the present argument does not involve any kind of underlying time reversal. Here the time reversal symmetry arises only at the statistical level of description
represented by the Markov evolution ruled by the transition rates. 

Notice also  that, as only certain ratios are fixed, different ergodic
deterministic microscopic dynamics can lead to very different transition rates,
a remnant of the fact that the coupling between a large part and the small part
can take any value a priori.

Formula \eqref{ergodic} is also a clue to understand a contrario what kind of
physical input is needed for the homogeneous Markov approximation to be valid.
Indeed, why didn't we do the homogeneous Markov approximation directly on the
small part ? We could certainly imagine dynamics making this a valid choice.
However, it is in general incompatible with the pattern of interactions (see the
end of subsubsection \ref{ErgodicityDiscrete}) which is
the basis of our argument. For instance, if in the coarse-graining procedure we had retained only the configurations $\C$'s of the small part, then the corresponding graph introduced in Appendix \ref{CoarseGrainedProperty}  would have been a cycle instead of a tree in the case of a small part made of two spins (see Ref.\cite{CornuBauerBETA}), and the crucial property \eqref{NumbersEqual} would have been  lost : over the period of $N$ time steps of the microscopic dynamics $N_{\C,\C'}\not=N_{\C',\C}$. 
In fact, we may expect, or impose on physical
grounds, that $\Omega_a$ will be exponentially large in the size of  large
part ${\cal B}_a$ (i.e.  its number of degrees of freedom $\N_a$), so that even the ratio
$\Omega_a(E'_a)/\Omega_a(E_a)=\Omega_a(E_a-\left[\En(\C') -\En(\C)\right])/
\Omega_a(E_a)$ will vary significantly over the trajectory, meaning that
transition probabilities involving only the small part  cannot be taken to be constant
along the trajectory: the energies of the large parts are relevant variables.

\subsection{Transient regime  when large parts are described in the thermodynamic limit}
\label{TransientSec}

\subsubsection{Large parts in the thermodynamic limit}

We now assume that the large parts are large enough that they are accurately
described by a thermodynamic limit, which we take at the most naive level. To
recall what we mean by that, we concentrate on one large part for a while, and
suppress the index used to label it. Suppose this large part has $\N$  degrees of freedom, and suppose that energies are close to an energy $E$ for which the Boltzmann
entropy is $\SB(E)$. That the thermodynamic limit exists means that if one lets
$\N \rightarrow +\infty$ while the ratio $E/\N$ goes to a finite limit
$\epsilon$, there is a differentiable function $s^B(\epsilon)$ such that the
ratio $\SB(E)/\N$ goes to $s^B(\epsilon)$. The quantity 
\be
\label{defbeta}
\frac{ds^B}{d\epsilon}\equiv \beta
\ee
is nothing but the inverse temperature. In that case, as long as $\Delta E \ll E$
 (where $E$ scales as $\N\Escale$ with $\Escale$ some finite energy scale), $\SB(E+\Delta E)-\SB(E) \rightarrow \beta \Delta E$ when $\N \rightarrow
+\infty$. For $\N$ large enough, the relation $\SB(E+\Delta E)-\SB(E) \sim \beta
\Delta E$ is a good approximation.

\subsubsection{Transient regime  and local  detailed balance (LDB )}

Notice that when transitions occur, which, by the definition of $\tau$ in \eqref{TauxMesoW0}, happens typically once  
on the macroscopic time scale, the changes in the
energies of the large parts are finite, so that over long windows of time
evolution, involving many changes in the small part,  the relation 
\be
\label{DifferenceSB}
\SB_a(E'_a)-\SB_a(E_a) \sim \beta_a
\left[E'_a-E_a\right]
\ee
is not spoiled, where $E'_a$ and $E_a$ are the energies in large part ${\cal B}_a$ at
any moment within the window.

In fact the larger the large parts, the longer
the time window for which \eqref{DifferenceSB} remains valid. The relation between the sizes 
$\N_a$'s  of the large parts and of the length of the
time window depends on the details of $\Udisc$ (which still has to fulfill the
imposed physical conditions). This relation also depends on the values of the energy per degree of freedom in every  large part,
$E_a/\N_a$,  which are essentially constant in such a window.

Because of the ergodicity hypothesis, the largest window (of size
comparable to the period of $\Udisc$ to logarithmic precision) has the property that the
energies $E_a$'s in the large parts will be such that all $\beta_a$'s are close to
each other and  the system will be at equilibrium. Indeed,
inside the largest time window, the
system remains in the region of the energy level where the
energies $E_a$'s   are the most probable, and in the  thermodynamic limit the most probable values for the $E_a$'s in the microcanonical ensemble are the values $E_a^\star$'s  that maximize the product $\prod_{a}\Omega_a(E_a)$ under the constraint $E=\sum_aE_a$ (since the system energies are negligible with respect to those of the large parts). The latter  maximization condition is equivalent to the equalities $ds^B/d\epsilon_a(\epsilon_a^\star)=ds^B/d\epsilon_b(\epsilon_b^\star)$
for all pairs of large parts.

However, if the system
starts in a configuration such that the $\beta_a$'s are distinct, the time window
over which $E_a/\N_a$ and $\beta_a$ are constant (to a good approximation)
will be short with respect to the period of the microscopic dynamics, but long enough that \eqref{DifferenceSB} still holds for a long time  interval.
Then by putting together
the information  on the ratio of transition rates  in terms of  Boltzmann entropies  \eqref{ergodic}, the transient regime  approximation \eqref{DifferenceSB} and  the energy conservation \eqref{conserE},  we get
\be
\label{ergodic2}
\frac{W(\C',\underline{E}' \overset{\F_a}{\leftarrow}
  \C,\underline{E})}{W( \C,\underline{E} \overset{\F_a}{\leftarrow}
  \C',\underline{E}')}\sim e^{-\beta_a\left[\En(\C')-\En(\C)\right]}.  
\ee 
Now the right-hand side depends only on the configurations of the small system,
and the parameters $\beta_a$'s are constants. 
Letting the large parts get larger and larger while adjusting the physical
properties adequately, we can ensure that the
time over which \eqref{ergodic2} remains valid gets longer and longer, so, in
the thermodynamic limit for the large parts, the transient regime  lasts
forever. This situation is our main interest in what follows.

In the transient regime the transition rate from configuration $\C$ to configuration $\C'$ is denoted as  
$(\C'\vert \Trans\vert \C)$.
As well as the transition rates
$W(\C',\underline{E}'\overset{\F_a}{\leftarrow}\C,\underline{E})$  the rates $(\C'\vert \Trans\vert \C)$ must satisfy the three consequences derived from the properties of the underlying microscopic deterministic dynamics pointed out in subsubsection \ref{ErgodicityDiscrete}, namely ergodicity,  energy conservation and specific interaction pattern.
 First, as shown in subsubsection  \ref{Microcansubsubsection} for the mesoscopic configurations $(\C, \underline{E})$,  any configuration $\C$ can be reached by a succession of jumps with
non-zero transition rates from any configuration $\C'$, namely in the network
representation of the stochastic evolution 
\be
\label{Mirreducible0}
\quad\textit{the graph  associated with the transition rates is connected.}\quad
\ee
In other words the Markov matrix  defined from the transition rates must  be irreducible and the property \eqref{Mirreducible0} is referred  to as the \textit{irreducibility} condition. Second, from \eqref{ergodicRev} the transition rates must obey
the \textit{microscopic reversibility} condition 
 for any couple of configurations $(\C,\C')$, namely
\be
\label{MicroRevCond0}
(\C'\vert \Trans \vert \C)\not=0 \qquad \Leftrightarrow \qquad (\C\vert \Trans
\vert \C')\not=0.
\ee 
Third,
from \eqref{ergodic2} one gets 
a constraint  obeyed by
the ratio of transition rates in the transient regime,
\be
\label{ergodicThermo}
\textrm{for $\C'\in\F_a(\C)$}\quad
\frac{(\C'\vert \Trans \vert \C)}
{(\C\vert \Trans \vert \C')}=e^{-\beta_a\left[\En(\C')-\En(\C)\right]}.
\ee
The latter relation is the  so-called local  detailed balance (LDB), which is also referred to in the literature as the ``generalized detailed balance''.

We stress that, by selecting a time window while taking the thermodynamic limit for  the large parts, the microcanonical detailed balance \eqref{ratioWmc} is replaced by the local  detailed balance \eqref{ergodic}, except in the case of the largest time window where all $\beta_a$'s are equal. In the latter case, the microcanonical detailed balance \eqref{ratioWmc} is replaced by the canonical detailed balance and the statistical time reversal symmetry is preserved. Indeed, the equilibrium thermodynamic  regime is reached either if the we start from a situation in
which $\prod_a \Omega_a(E_a)$ is close to its maximum along the trajectory in the energy level $E_\text{dec}=E$, or if we 
wait long enough so that $\prod_a \Omega_a(E_a)$ becomes close to this maximum.
As recalled above, this is true for most of the period of the microscopic dynamics, but reaching this
situation may however take a huge number of time steps if the starting point was
far from the maximum.
By an argument similar to that used in the derivation of \eqref{DifferenceSB}, when $\prod_a \Omega_a(E_a)$ is closed to its maximum and the large parts are considered in the thermodynamic limit, all $\beta_a$'s are equal to the same value $\beta$ and  the relative weight of two configurations in the microcanonical ensemble, $\Probmc(\C',\underline{E}')/\Probmc(\C,\underline{E})$ given by \eqref{Probmicro}, tends to $\exp\left(-\beta[\En(\C')-\En(\C)]\right)$. 
Then the equilibrium microcanonical distribution $\Probmc(\C,\underline{E})$  tends to the canonical distribution   
\be
\label{expProbcan}
\Probcan^{\beta}(\C)\equiv\frac{e^{-\beta \En(\C)}}{Z(\beta)},
\ee
where $Z(\beta)$ is the canonical partition function at the inverse temperature $\beta$. 
Meanwhile, 
the detailed balance relation \eqref{ratioWmc} in the
microcanonical equilibrium ensemble for the transition rates
$W(\C',\underline{E}'\leftarrow\C,\underline{E})$ 
becomes a detailed balance relation in the canonical ensemble
at the inverse temperature $\beta$ of the whole system for the transition rates $(\C'\vert
\Trans\vert \C)$, namely
\be
\label{DetailedBalanceCan}
\frac{(\C'\vert \Trans \vert \C)}
{(\C\vert \Trans \vert \C')}=e^{-\beta\left[\En(\C')-\En(\C)\right]}
=\frac{\Probcan^{\beta}(\C')}{\Probcan^{\beta}(\C)}.
\ee
The local  detailed balance \eqref{ergodicThermo}, valid in transient regimes, differs from the  latter detailed balance  in the canonical ensemble  by two features : the various $\beta_a$'s of the distinct large parts appear in place of the
common equilibrium inverse temperature $\beta$, and the stationary distribution for the transition rates is not known a priori.

We conclude this discussion with the following remarks. We have not tried to
exhibit explicit physical descriptions of the large parts, or explicit
formul\ae\ for the dynamical map $\Udisc$. Though it is not too difficult to give
examples for fixed sizes of the large parts, it is harder to get a family of
such descriptions sharing identical physical properties for varying large part
sizes, a feature which is crucial to really make sense of the limits we took
blindly in our derivation. It is certainly doable, but
cumbersome, and we have not tried to pursue this idea.
Let us note also that in
principle, taking large parts of increasing sizes can be used to enhance  the
validity of the approximation of the (discrete-time) Markov chain by a
(continuous-time) Markov process. As the physics of the continuous time limit
does not seem to be related to the physics of convergence towards a heat bath
description we have preferred to keep the discussion separate, taking a
continuous-time description as starting point.

\subsubsection{Expression of LDB  in terms of exchange entropy variation}

Observe that though we have given no detailed analysis of the dynamics or the
statistical properties of the large parts, their influence on the effective
Markov dynamics of the small system enters only through the inverse temperatures
$\beta_a$ defined in \eqref{defbeta}.  So we can consistently assume that each
large part becomes a thermal bath with its own temperature. The leading term in
$\SB_a(E'_a)-\SB_a(E_a)$ is the variation $\delta S_a^{\scriptscriptstyle TH}
(\C'\leftarrow \C)$ of the thermodynamic entropy of bath ${\cal B}_a$ when it flips the small
system from configuration $\C$ to configuration $\C'$,
\be
\label{defdeltaStherm}
\delta S_a^{\scriptscriptstyle TH}
(\C'\leftarrow \C)=
\begin{cases}
\beta_a \left[E'_a-E_a\right]  &\textrm{if}\quad  \C'\in\F_a(\C) \\
0 &\textrm{otherwise}.
\end{cases}
\ee 
Then we have an idealized description of a thermal contact
between  heat baths.  This is the situation on which we concentrate in this
paper.
 
By definition of a heat source, the variation $\delta S_a^{\scriptscriptstyle TH}(\C'\leftarrow \C)$ of the thermodynamic entropy of bath ${\cal B}_a$ when it flips the system from configuration $\C$ to configuration $\C'$  reads
\be
\label{deltaSatherm}
\delta \STH_a
(\C'\leftarrow \C)=- \beta_a \delta\Heatm_a(\C'\leftarrow\, \C),
\ee
where $\delta\Heatm_a(\C'\leftarrow \C)$ is the heat received by the small system from part ${\cal B}_a$. According to the expression \eqref{defdeltaStherm} and to the energy conservation relation \eqref{conserE},
\be
\label{defdeltaq}
\begin{cases}
\delta\Heatm_a(\C'\leftarrow\, \C)=\En(\C')-\En(\C)  &\textrm{if}\quad  \C'\in\F_a(\C) \\
 \delta\Heatm_a(\C'\leftarrow\, \C)=0 &\textrm{otherwise}.
  \end{cases}
\ee

  Let us introduce $\deltaexch S(\C'\leftarrow\, \C
)$ the exchange entropy variation of the small system (see footnote \ref{exchangename})  that  is associated with the heat exchanges  with the  thermostats when the small system goes from configuration $\C$ to configuration $\C'$. Thanks to the definition \eqref{defdeltaStherm}
\be
\deltaexch S(\C'\leftarrow\, \C)
\equiv
-\sum_a\delta \STH_a(\C'\leftarrow\, \C).
\ee
Then the local  detailed balance \eqref{ergodicThermo} can be rewritten  in a form which does not involve explicitly the heat bath responsible for the transition from $\C$ to $\C'$,
\be
\label{MDBexch}
\frac{(\C'\vert \Trans \vert C)}{(\C\vert \Trans \vert C')}=e^{-\deltaexch S(\C'\leftarrow\, \C)}.
\ee

\section{Exchange entropy variation and symmetries at finite time under LDB }
\label{FiniteTimeSymmetry}

\subsection{LDB  and symmetry between time-reversed histories}
\label{SymmetryHistories}

For a history $\Hist$ where the system starts in configuration $\C_0$ at time $t_0=0$ and  ends in configuration $\C_f$ at time $t$ after going through successive configurations $\C_0$, $\C_1$,\ldots, $\C_N=\C_f$, the exchange entropy variation $\Deltaexch S[\Hist]$  corresponding to the history is defined from the heat amounts $\Heat_1[\Hist]$ and $\Heat_2[\Hist]$ received from two  thermal baths as
\be
\label{DefSexch}
\Deltaexch S[\Hist]\equiv\beta_1 \Heat_1[\Hist]+\beta_2 \Heat_2[\Hist]
\quad\textrm{with}\quad  \Heat_a[\Hist]\equiv\sum_{i=0}^{N -1}\delta \Heatm_a(\C_{i+1}\leftarrow\C_i).
\ee

 The $N$ instantaneous jumps from one configuration to another occur at $N$ successive intermediate times $T_i$ which are continuous stochastic variables : the system jumps from $\C_{i-1}$ to $\C_i$ at time $T_i$ ($i=1,\ldots,N$) in the time interval $[t_i, t_i+dt_i[$, with $t_0<t_1<t_2<\cdots<t_N<t$. The probability measure for such  a history   is related to the probability density $\Probdist_{\C_f,\C_0}\left[\Hist \right]$ by
\be
\label{defProbdisHist}
d\Prob_{\C_f, \C_0}\left[\Hist \right]
\equiv dt_1\ldots dt_N \Probdist_{\C_f,\C_0}\left[\Hist \right]
\ee
where, for a time-translational invariant process,  
\bea
\Probdist_{\C_f,\C_0}\left[\Hist \right]&=&e^{-(t-t_N)  \Lambda(\C_N)}(\C_N\vert \Trans \vert \C_{N-1}) e^{-(t_N-t_{N-1})\Lambda(\C_{N-1})}
\\
\nonumber
&&\qquad\qquad\qquad\qquad \times
  \cdots
 e^{-(t_2-t_1)\Lambda(\C_1)}(\C_1\vert \Trans \vert \C_0) e^{-(t_1-t_0) \Lambda(\C_0)}
\eea
and $\Lambda(\C)$ is  the total exit rate (also called escape rate) from configuration $\C$,
$\Lambda(\C)\equiv\sum_{\C'\not=\C} (\C'\vert \Trans \vert \C)$.
The average of a functional $F[\Hist]$ over the histories that start in configuration $\C_0$ and end in configuration $\C_f$    is computed as $\Esp{F}_{\C_f,\C_0}=\int d\Prob_{\C_f, \C_0}\left[\Hist \right] F[\Hist]$
with
\be
\label{defintdProb}
\int d\Prob_{\C_f, \C_0}\left[\Hist \right] =\sum_{N=0}^{+\infty}\sum_{\C_1}\ldots \sum_{\C_{N-1}} 
\int_{t_0<t_1<\ldots< t_N}
d\Prob_{\C_f, \C_0}\left[\Hist \right].
\ee
Then the average of a functional when the initial distribution of configurations is $\Prob_0$ reads
\be
\label{defEspHist}
\Esp{F}_{\Prob_0}=\sum_{\C_f} \sum_{\C_0}\Prob_0(\C_0)\int d\Prob_{\C_f, \C_0}\left[\Hist \right]  F[\Hist].
\ee
Let  $\TimeRev$ be the time reversal operator for histories.
 If $\Hist$ is a history that starts at time $t_0=0$ in $\C_0$ and ends  at time $t$ in $\C_f$ after $N$ jumps from $\C_{i-1}$ to  $\C_i$ at time $T_i$,
$\TimeRev\Hist$ is a history that starts at time $t_0=0$ in $\C_f$ and ends at $\C_0$ at time $t$ after $N$  jumps from $\C'_{i-1}$ to  $\C'_i$ at time $T'_i$  with $\C'_i=\C_{N-i}$ and  $T'_i= t-T_{N-i+1}$ , namely
\bea
\Hist:\qquad\C_0\,\textrm{at}\, t_0=0 \quad 
&&\C_0 \overset{T_1}{\longrightarrow}\C_1\cdots  \C_{N-1}\overset{T_{N}}{\longrightarrow}\C_f
\\
\nonumber
\TimeRev\Hist:\qquad
\C_f\,\textrm{at}\, t_0=0 \quad
&&\C_f \overset{T'_1}{\longrightarrow}\C_{N-1}\cdots  \C_{1}\overset{T'_{N}}{\longrightarrow}\C_0.
\eea
From the definition \eqref{defProbdisHist} of the measure $d\Prob_{\C_0, \C_f} $ over histories starting in configuration $\C_0$ and ending in configuration $\C_f$,
\be
\label{defratio}
\frac{d\Prob_{\C_f, \C_0}\left[\Hist \right] }{d\Prob_{\C_0, \C_f} \left[\TimeRev\Hist \right]}=
 \prod_{i=0}^{N-1}\frac{(\C_{i+1}\vert \Trans \vert \C_i)}{(\C_{i}\vert \Trans \vert \C_{i+1})}.
\ee

When the transition rates obey the local  detailed balance \eqref{MDBexch} written in terms of $\deltaexch S(\C'\leftarrow\,\C)$, the exchange entropy variation for the history, defined in \eqref{DefSexch}, can be rewritten as 
\be
\label{SexchActionF}
\Deltaexch S[\Hist]
\underset{LDB}{=}-\ln \prod_{i=0}^{N-1}\frac{(\C_{i+1}\vert \Trans \vert \C_i)}{(\C_{i}\vert \Trans \vert \C_{i+1})},
\ee
and equation \eqref{defratio} can be rewritten as
\be
\label{historypropertySexch}
\frac{d\Prob_{\C_f, \C_0}\left[\Hist \right] }{d\Prob_{\C_0, \C_f} \left[\TimeRev\Hist \right]}
\underset{LDB}{=}e^{-\Deltaexch S[\Hist]}.
\ee
We stress that, according to \eqref{SexchActionF},   when the LDB is satisfied  the expression   of the exchange entropy variation for a history defined in \eqref{DefSexch}  coincides with the opposite of
 the ``action functional'' introduced by Lebowitz and Spohn in Ref.\cite{LebowitzSpohn1999} in the generic case where the LDB  does not necessarily hold.

\subsection{Symmetry between time-reversed evolutions with fixed heat amounts}
\label{SymmetryConfigMesoSec}

The probability $\Prob\left(\C_f \vert \Heat_1,\Heat_2, t\vert  \C_0\right)$ that the system has evolved from  configuration $\C_0$ at $t_0=0$ to  configuration $\C_f$ at $t$ while receiving the heat amounts $\Heat_1$ and $\Heat_2$  from the thermostats 1 and 2 reads
\be
\Prob\left(\C_f \vert \Heat_1,\Heat_2; t\vert  \C_0\right)\equiv
\int d\Prob_{\C_f, \C_0}\left[\Hist \right]\delta\left( \Heat_1[\Hist]-\Heat_1\right)\delta\left( \Heat_2[\Hist]-\Heat_2\right),
\ee
where $\int d\Prob_{\C_f, \C_0}$ denotes the ``summation'' over the histories from $\C_0$ to $\C_f$ defined in \eqref{defintdProb}.
The time-reversal symmetry property \eqref{historypropertySexch} for the history measure $d\Prob_{\C_f, \C_0}\left[\Hist \right]$ implies the  following relation between probabilities of forward and backward evolutions where initial and final configurations are exchanged (and  heat amounts are changed into their opposite values),
\be
\label{TimeRevPQ1PQ2CfC0}
\frac{\Prob\left(\C_f \vert \Heat_1,\Heat_2; t\vert  \C_0\right)}{\Prob\left(\C_0\vert -\Heat_1,-\Heat_2; t\vert \C_f\right)}
=e^{-\Deltaexch S(\Heat_1,\Heat_2)},
\ee
with the definition
\be
\label{defDeltaexchHeat}
\Deltaexch S(\Heat_1,\Heat_2)\equiv \beta_1\Heat_1+\beta_2\Heat_2.
\ee
An analogous relation for $\Deltaexch S$ in place of $(\Heat_1,\Heat_2)$ is derived in \cite{Jarzynski2000} in the case where the microscopic dynamics of the heat baths is assumed to be Hamiltonian.

\subsection{Symmetries in protocols starting from an equilibrium state}
\label{SymProtocoleEqSec}

We consider a protocol where the system is prepared in an equilibrium state  at the inverse temperature $\beta_0$ and suddenly put at time $t_0=0$ in thermal contact  with the two thermostats at the inverse temperatures $\beta_1$ and $\beta_2$ respectively. Then the system evolution is a relaxation from an equilibrium state to a stationary non-equilibrium state.

The initial  equilibrium distribution  at the inverse temperature $\beta_0$ is the canonical  distribution \eqref{expProbcan}. $Z(\beta_0)$ cancels in the ratio $\Probcan^{\beta_0}(\C_0)/\Probcan^{\beta_0}(\C_f)$ and
\be
 \label{lnProb}
\ln \frac{\Probcan^{\beta_0}(\C_0)}{\Probcan^{\beta_0}(\C_f)}=\beta_0\left[\En(\C_f)-\En(\C_0)\right]=\beta_0 (\Heat_1+\Heat_2),
\ee
where the last equality is enforced by energy conservation. 
Then the time-reversal symmetry \eqref{TimeRevPQ1PQ2CfC0} and the specific form
\eqref{lnProb} for $\ln[\Probcan^{\beta_0}(\C_0)/\Probcan^{\beta_0}(\C_f)]$   imply that
\be
\label{TimeRevPQ1PQ2CfC0Eq}
\frac{\Prob\left(\C_f \vert \Heat_1,\Heat_2; t\vert  \C_0\right)\Probcan^{\beta_0}(\C_0)}{\Prob\left(\C_0\vert -\Heat_1,-\Heat_2; t\vert \C_f\right)\Probcan^{\beta_0}(\C_f)}
=e^{-\Deltaexchexcess S(\Heat_1,\Heat_2)},
\ee
where the excess exchange entropy variation $\Deltaexchexcess S(\Heat_1,\Heat_2)$  is defined as the difference between the  exchange entropy variation in an evolution under the non-equilibrium constraint  $\beta_1\neq\beta_2$ where the system receives heat amounts $\Heat_1$ and $\Heat_2$ and that in an evolution under the equilibrium condition $\beta_1=\beta_2=\beta_0$ where the system would received the same heat amounts. It reads
\be
\label{expDeltaexchexcess}
\Deltaexchexcess S(\Heat_1,\Heat_2)=\Deltaexch S(\Heat_1,\Heat_2)-\beta_0(\Heat_1+\Heat_2)=
(\beta_1-\beta_0)\Heat_1+(\beta_2-\beta_0)\Heat_2.
\ee
A crucial point is that $\Deltaexchexcess S(\Heat_1,\Heat_2)$ does not depend explicitly on the  initial and final configurations and is only a function of the heat amounts received from the thermal baths,

As a consequence, the measurable joint distribution $\Prob_{\Probcan^{\beta_0}}\left(\Heat_1,\Heat_2; t\right)$ for the heat amounts $\Heat_1$ and $\Heat_2$ received between $t_0=0$ and $t$ when the initial configuration of the system is distributed according to $\Probcan^{\beta_0}$, namely
$\Prob_{\Probcan^{\beta_0}}\left(\Heat_1,\Heat_2; t\right)=\sum_{\C_0,\C_f}\Prob\left(\C_f \vert \Heat_1,\Heat_2, t\vert  \C_0\right)\Prob_{\Probcan^{\beta_0}}(\C_0)$, satisfies the identity
\be
\label{ratioProbQ1Q2}
\frac{\Prob_{\Probcan^{\beta_0}}\left(\Heat_1,\Heat_2; t\right)}{\Prob_{\Probcan^{\beta_0}}
\left(-\Heat_1,-\Heat_2; t\right)}=e^{-\Deltaexchexcess S(\Heat_1,\Heat_2)}.
\ee
Subsequently the  measurable quantity $\Deltaexchexcess S(\Heat_1,\Heat_2)$, with the distribution probability
\\
$\Prob_{\Probcan^{\beta_0}}\left(\Deltaexchexcess S\right)=\sum_{\Heat_1,\Heat_2}
\delta\left(\Deltaexchexcess S-(\beta_1-\beta_0)\Heat_1-(\beta_2-\beta_0)\Heat_2\right)
\Prob_{\Probcan^{\beta_0}}\left(\Heat_1,\Heat_2; t\right)$ 
obeys  the symmetry relation at any finite time, which may be referred to as a detailed fluctuation relation,
\be
\label{DFRDeltaexchexcess}
\frac{\Prob_{\Probcan^{\beta_0}}\left(\Deltaexchexcess S\right)}{\Prob_{\Probcan^{\beta_0}}\left(-\Deltaexchexcess S\right)}=e^{-\Deltaexchexcess S}.
\ee
The latter relation itself entails the identity, which may be referred to as an integral fluctuation relation,
\be
\label{IFRDeltaexchexcess}
\Esp{e^{\Deltaexchexcess S}}_{\Probcan^{\beta_0}}=1.
\ee
By using Jensen's inequality we get the inequality $-\Esp{\Deltaexchexcess S}\geq 0$.
To our knowledge the two relations \eqref{DFRDeltaexchexcess} and \eqref{IFRDeltaexchexcess}  have not appeared explicitly in the literature, though the calculations involved in the derivation of the present finite-time fluctuation relations are analogous to those that lead to finite-time detailed fluctuation relations for protocols where the system is in thermal contact with only one heat bath and is driven out of equilibrium by a time-dependent external parameter (see the argument first exhibited by Crooks \cite{Crooks1999} for work fluctuations and then Seifert \cite{Seifert2005} for the entropy production along a stochastic trajectory (see also  the review \cite{Seifert2008})). 
The latter class of protocols is very different as for the physical mechanisms that they involve : the changes in energy level populations are caused by energy exchanges with only one thermal bath and the system is driven out of equilibrium by the time dependence of the energy levels enforced by external time-dependent forces  \cite{Crooks1999Thesis}. 
Moreover, in Jarzynski-like protocols the system evolves from an initial equilibrium state and measurements are performed until work ceases to be provided to the system, which then relaxes to another equilibrium state.
The present protocol does not either involve the comparison of forward and backward  evolutions corresponding to  two different series of experiments, as needed for Crooks relation. (In Crooks' argument the initial  configurations for the forward and backward evolutions are distributed with different equilibrium probabilities, $\Probcan^{\beta_0}$ and $\Probcan^{\beta_f}$, whereas forward and backward evolutions with the same initial distribution had already been considered in 
\cite{BochkovKuzovlev1981a,BochkovKuzovlev1981b}). 
In Hatano-Sasa-like protocols  the system evolves from an initial non-equilibrium steady state to another one (and then housekeeping heats and excess heats are introduced as in the steady state thermodynamics introduced by Oono and Paniconi \cite{OonoPaniconi1998}). In the  finite-time protocol considered here the system starts in an equilibrium state and at time $t$ it has not yet  reached the steady state controlled by $\beta_1$ and $\beta_2$. Moreover, the integral fluctuation relation \eqref{IFRDeltaexchexcess}  differs from the Hatano-Sasa relation \cite{HatanoSasa2001} in the sense that the quantity to be averaged over repeated experiments does not involve the probability distribution of the system.

\subsection{Symmetries in protocols starting from a stationary state with a canonical distribution}
\label{SymProtocolNESSSec}

For some systems, such as the two-spin model studied in Ref.\cite{CornuBauerBETA}, the stationary distribution 
when the thermostats are at the inverse temperatures $\beta_1$ and $\beta_2$ proves to be a canonical distribution at the effective inverse temperature $\betaeff(\beta_1,\beta_2)$. 

When the system is  prepared in a stationary state  between two heat baths at the inverse temperatures $\beta_1^0$ and $\beta_2^0$ and then put in thermal contact with two thermostats at the inverse temperatures $\beta_1$ and $\beta_2$ at time $t_0=0$,
 the protocol describes the relaxation from a given stationary state corresponding to $(\beta^0_1,\beta_2^0)$ to another stationary state
corresponding to $(\beta_1,\beta_2)$.
When the initial stationary state has  the canonical  distribution at the effective inverse temperature $\beta_{\star}^0=\betaeff(\beta_1^0,\beta_2^0)$, the argument of the previous subsection can be repeated and the equalities \eqref{DFRDeltaexchexcess} and \eqref{IFRDeltaexchexcess}  
 still hold with  $\beta_0$ replaced by $\beta_\star^0$ and $\Deltaexchexcess S$ replaced by
\be
\Delta_\text{\mdseries exch}^{\text{\mdseries excs},\beta_\star^0} S(\Heat_1,\Heat_2)=(\beta_1-\beta_{\star}^0)\Heat_1+(\beta_2-\beta_{\star}^0)\Heat_2.
\ee

When the system is  already  in the stationary state  corresponding to  the inverse temperatures $\beta_1$ and $\beta_2$ at time $t_0=0$,  the equalities \eqref{DFRDeltaexchexcess} and \eqref{IFRDeltaexchexcess} for $\Deltaexchexcess S$ still hold with $\beta_\star(\beta_1,\beta_2)$ in place of $\beta_0$ :
\be
\frac{\Probst\left(\Delta_\text{\mdseries exch}^{\text{\mdseries excs},\beta_\star} S\right)}{\Probst\left(-\Delta_\text{\mdseries exch}^{\text{\mdseries excs},\beta_\star} S \right)}=e^{-\Delta_\text{\mdseries exch}^{\text{\mdseries excs},\beta_\star} S},
\ee
where the subscript ``st'' in the notation for the probability is  a reminder of the fact that  the initial configurations are distributed according to the stationary measure, which is equal to
$\Probcan^{\betaeff}$ in the present case. Another detailed fluctuation relation involving the forward histories for the original dynamics and the backward histories for the dual reversed dynamics is 
derived in \cite{EspositoVanDenBroeck2010PRL} for the case where the external parameters also vary during the time interval $]t_0,t]$;   these considerations are out of the scope of the present paper.

\section{Extension of previous results to a larger class of models}
\label{GeneraliztionSeveralCurrents}

\subsection{Generic expression of exchange entropy variation}
\label{GenericExchS}

In this section we consider the generic case where   the finite-size system $\syst$ is made of $\nu_s$ species of 
mobile elementary constituents   and  can occupy a domain  whose  boundaries may be mobile interfaces. The degrees of freedom of every elementary constituent involve the site where it sits in discretized space and  some possible internal degrees of freedom. In the following we call degrees of freedom of a configuration $\C$ of the system $\syst$  the degrees of freedom of the  elementary constituents in this configuration. (The number of constituents of every species may vary from one configuration to another.) When  some boundaries are mobile interfaces, the description of any configuration $\C$ of the system not only involves the values of its degrees of freedom  
but it is also specified by the positions of the  interfaces that surround the domain, called ${\cal D}(\C)$, that the system $\syst$  can  occupy. For each configuration $\C$ one can  define the following global quantities : the energy $\En(\C)$, the volume $v(\C)$ of ${\cal D}(\C)$ and the total number of elementary constituents $n(\C)=\sum_{s=1}^{\nu_s}n_s(\C)$, where $n_s(\C)$ is the  number of elementary constituents of species $s$ that sit in ${\cal D}(\C)$. All these quantities are assumed to take a finite number of values. The system $\syst$ is in contact with several macroscopic bodies ${\cal B}_a$'s.

A crucial assumption is that in the course of the ergodic deterministic microscopic  dynamics of the whole system ($\syst$ and the large parts ${\cal B}_a$), for a given configuration $\C$, the domain ${\cal D}(\C)$ can be divided in several disjoint subdomains ${\cal D}_a(\C)$'s such that some  boundary portion of ${\cal D}_a(\C)$ can move only thanks to a corresponding volume variation of large part ${\cal B}_a$  and  the values of the degrees of freedom that sit inside ${\cal D}_a(\C)$ can vary only by  exchanging microscopically conserved  quantities (energy  and/or matter)  with  the corresponding large part ${\cal B}_a$. Then  a   jump of  system $\syst$  from  configuration $\C$  to another one $\C'$  is allowed only if ${\cal D}(\C)$ and ${\cal D}(\C')$ differ by a displacement of some  boundary portion of only one ${\cal D}_a(\C)$ and by different values of the degrees of freedom inside ${\cal D}_a(\C)$ and ${\cal D}_a(\C')$ ; then we use the notation $\C'\in \F_{a}(\C)$. 
Moreover  the corresponding jump of the microscopic configuration of ${\cal B}_a$ is such that   conservation rules hold for the sum  of the energies of $\syst$ and ${\cal B}_a$, $\En(\C')+E'_a=\En(\C)+E_a$, for the sum of the volumes that they occupy, 
$v(\C')+V'_a=v(\C)+V_a$, and for the sum of the numbers of elementary constituents of species $s$ that they contain, $n_s(\C')+N'_{a,s}=n_s(\C)+N_{a,s}$, where $E_a$, $V_a$ and the $N_{a,s}$ are the values of the extensive parameters that characterize body ${\cal B}_a$ (and the prime denotes their values after the configuration jump). For instance if   system  $\syst$ is a mobile  diathermal thin solid wall separating a vessel in two parts filled with gases kept at different temperatures and pressures, then $\syst$ can be viewed as made of two layers of constituents, each of which interacts respectively  with body
${\cal B}_1$ and  ${\cal B}_2$. Then an infinitesimal displacement of $\syst$ such that the volume of ${\cal B}_1$ increases while that of ${\cal B}_2$ decreases by the same absolute amount can be  decomposed into two microscopic configuration jumps of the global system :  in the first (second) jump only the layer in contact with ${\cal B}_2$ (${\cal B}_1$) moves and the volume of $\syst$ increases (decreases), and after two  jumps the volume of $\syst$  has retrieved its initial value. 

Then the mesoscopic Markovian dynamics defined according to the prescription of subsection \ref{MarkovianApproximationConseq} is such that, when energy, volume and matter are exchanged, the transition rates obey the microcanonical  detailed balance
 \be
\label{ratioWmcgen}
 \frac{W(\C',E'_a,V'_a, \{N'_{a,s}\}\leftarrow \C,E_a,V_a, \{N_{a,s}\})}
  {W(\C,E_a,V_a, \{N_{a,s}\}\leftarrow \C',E'_a,V'_a, \{N'_{a,s}\})}=
\frac{\Omega_a(E'_a,V'_a, \{N'_{a,s}\})}
{\Omega_a(E_a,V_a, \{N_{a,s}\})}
\ee
with
\be
\label{ratioOmegaPattern}
\frac{\Omega_a(E'_a,V'_a, \{N'_{a,s}\})}
{\Omega_a(E_a,V_a, \{N_{a,s}\})}
\equiv e^{\SB_a(E'_a,V'_a, \{N'_{a,s}\})-\SB_a(E_a,V_a, \{N_{a,s}\})},
 \ee
where $\Omega_a(E_a,V_a, \{N_{a,s}\})$ is the number of configurations (or microstates) of large part ${\cal B}_a$ when it is isolated. 
In the following, in the spirit of the notations used by Callen \cite{Callen1960} we denote by $X_i^{(a)}$ the extensive macroscopic parameters of ${\cal B}_a$, namely in the generic case
\be
\label{examplesX}
X^{(a)}_0\equiv E_a, \quad  X^{(a)}_1\equiv V_a, \quad X^{(a)}_{1+s}=N_{a,s}.
\ee
The total number of elementary constituents in ${\cal B}_a$ is $N_a=\sum_sN_{a,s}$. The microscopic conservation rules that are associated with \eqref{ratioWmcgen} read
\be
\label{ConservationRuleGen}
X'^{(a)}_i-X^{(a)}_i=-\left[x_i(\C')-x_i(\C)\right] \quad   \text{if}\quad \C'\in\F_a(\C),
\ee
 with the same notations for  system $\syst$ as those introduced in \eqref{examplesX} for the macroscopic bodies.

If the  bodies ${\cal B}_a$ are so large that they can be  described by a thermodynamic limit, then, in a transient regime  where the macroscopic extensive parameters $X^{(a)}_i$'s have negligible relative variations, 
  ${\cal B}_a$ remains at thermodynamic equilibrium. 
The thermodynamic entropy  per elementary constituent, $\STH_a/N_a$, coincides with the thermodynamic limit of the Boltzmann entropy per  elementary constituent, $\SB_a/N_a$. The  intensive thermodynamic parameter $F^{(a)}_i$ conjugate to the extensive quantity $X_i^{(a)}$  by $F^{(a)}_i\equiv \partial \STH_a /\partial X^{(a)}_i$ (with $X^{(a)}_i$ defined in \eqref{examplesX}) is given by
\be
\label{examplesF}
F^{(a)}_0\equiv \beta_a, \quad  F^{(a)}_1\equiv \beta_a P_a, \quad F^{(a)}_{1+s}=-\beta_a \mu_{a,s},
\ee
where $\beta_a$ is the inverse thermodynamic temperature, $P_a$ the thermodynamic pressure and $\mu_{a,s}$ the chemical potential of species $s$ in ${\cal B}_a$. Then from the relations \eqref{ratioWmcgen} and \eqref{ratioOmegaPattern}, 
one can show, as in subsection \ref{TransientSec},  that in the  transient regime   the transition rates obey the local  detailed balance \eqref{MDBexch} where $\deltaexch S(\C'\leftarrow\, \C)$ is opposite to
 the infinitesimal variation at fixed intensive parameters of the thermodynamic entropy of the reservoir ${\cal B}_a$ that causes the jump   from $\C$ to $\C'$ under the conservation rules \eqref{ConservationRuleGen}. Therefore, if $\C'\in\F_a(\C)$
\be  
\label{ExplicitDeltaSexchGen}
 \deltaexch S(\C'\leftarrow\, \C)=\beta_a [\En(\C')-\En(\C)]
+\beta_a  P_a [v(\C')-v(\C)]- \beta_a\sum_s \mu_{a,s} [n_s(\C')-n_s(\C)].
\ee
In the case of pure thermal contact, the volume of  system $\syst$ does not vary, $v(\C)=v(\C')$, and there is no matter exchange ; then
   $\deltaexch S(\C'\leftarrow\, \C)=\beta_a [\En(\C')-\En(\C)]$ coincides with  $\beta_a$ times the opposite of the variation of the internal energy  of ${\cal B}_a$, which is equal in that case to the heat given by ${\cal B}_a$ at constant volume.
If system $\syst$ and macroscopic body ${\cal B}_a$ are compressible then, during energy exchanges such that the thermodynamic pressure $P_a$ of ${\cal B}_a$ remains fixed,   the variation of the internal energy of the macroscopic body ${\cal B}_a$ involves both heat and   pressure work ; then
$
\deltaexch S(\C'\leftarrow\, \C)=\beta_a [\En(\C')-\En(\C)]
+\beta_a  P_a [v(\C')-v(\C)]
$
 coincides with $\beta_a$ times the opposite of the variation of the enthalpy  of ${\cal B}_a$, which is equal in that case to the heat given by ${\cal B}_a$ at constant pressure.
 Moreover  system $\syst$ may  receive work from some conservative external forces $ f^\text{ext}_b$ (such as gravitational or electrical fields), each of which causes the variation of some global  coordinate $Z_b(\C)$ of the system (such as its mass center or its electrical  barycenter) but does not act upon the macroscopic bodies. In that case the mechanical energy $\En(\C)$ of system $\syst$ in configuration $\C$ is the sum of its internal energy  $\En_\text{int}(\C)$ and an external potential energy  $\En_\text{ext}(\C)$.   In all these situations the distances between the possible  energy levels $\En$ are time-independent, and  during the evolution  only the occupation of the energy levels  is modified, contrarily to the case where a time-dependent force acts on the system by changing the spacing between energy levels.

In order to handle compact notations for the successive variations  $\deltaexch S$ in the course of a history of system $\syst$, let us  introduce
the notation
$\delta  \chi_i^{(a)}(\C'\leftarrow\, \C)$ for the quantity with index $i$
received by the system from  reservoir ${\cal B}_a$ when the system jumps from $\C$ to
$\C'$, with the same convention as in definition \eqref{defdeltaq}, namely 
\be
\label{defdeltax}
\begin{cases}
\delta \chi_i^{(a)}(\C'\leftarrow\, \C)=x_i(\C')-x_i(\C)  &\textrm{if}\quad  \C'\in\F_a(\C) \\
\delta \chi_i^{(a)}(\C'\leftarrow\, \C)=0 &\textrm{otherwise}
  \end{cases}.
\ee
(In the case where $ \C'\in\F_a(\C)$ but where in fact reservoir ${\cal B}_a$ does not exchange quantity with index $i$ in the jump from $\C$ to $\C'$, $x_i(\C')-x_i(\C)=0$). Then the exchange entropy variation in a jump takes the form
\be
\label{ExplicitGenericdeltaexchS}
\deltaexch S(\C'\leftarrow\, \C)=\sum_{i}\sum_{a\in R(i)} F_i^{(a)}\delta \chi_i^{(a)}(\C'\leftarrow\, \C),
\ee 
where the first sum runs over the indices $i$ of the extensive quantities  defined in \eqref{examplesX}, the second sum runs  over the indices of the reservoirs that indeed can exchange quantity $i$ with the system and $R(i)$ denotes the set of the latter reservoirs.

\subsection{Consequences of LDB  at finite time}

In the form \eqref{MDBexch} where it involves the microscopic exchange entropy variation $\deltaexch S(\C'\leftarrow\, \C)$ associated with a  jump from configuration $\C$ to  configuration $\C'$, the local  detailed balance entails the symmetry \eqref{historypropertySexch} between the probabilities for time-reversed histories. At a more mesoscopic level,
  let us compare the probability of all evolutions from  configuration $\C_0$ to  configuration $\C_f$, in the course of which the system receives given cumulative quantities  with index $i$ ${\cal X}^{(a)}_i=\sum \delta \chi^{(a)}_i$ from each reservoir ${\cal B}_a$,  and the probability of the reversed evolutions, namely evolutions from $\C_f$ to $\C_0$ where the cumulative quantities are $-{\cal X}^{(a)}_i$'s.  The symmetry \eqref{TimeRevPQ1PQ2CfC0} written in the case of thermal contact takes the following form in the generic case
\be
\label{TimeRevPQ1PQ2CfC0Gen}
\frac{\Prob\left(\C_f \vert\{{\cal X}^{(a)}_i\}; t\vert  \C_0\right)}{\Prob\left(\C_0\vert \{-{\cal X}^{(a)}_i\}; t\vert \C_f\right)}
=e^{-\Deltaexch S(\{{\cal X}^{(a)}_i\})},
\ee
with, according to  \eqref{ExplicitGenericdeltaexchS},
\be 
\label{defgenDeltaexchS}
\Deltaexch S(\{{\cal X}^{(a)}_i\})=\sum_i\sum_{a\in R(i)} F^{(a)}_i{\cal X}^{(a)}_i.
\ee
In \eqref{TimeRevPQ1PQ2CfC0Gen} and in the formul\ae\, derived from it in the following, 
${\cal X}^{(a)}_i$ occurs only if reservoir ${\cal B}_a$ indeed exchanges quantity with index $i$.
The ratio of probabilities in \eqref{TimeRevPQ1PQ2CfC0Gen} does not  explicitly  depend on the initial and final configurations $\C_0$ and $\C_f$. There is only an implicit dependence on these configurations through the conservation rules that the  cumulative exchange quantities ${\cal X}^{(a)}_i$ for a given history must satisfy, namely $\sum_{a\in R(i)} {\cal X}^{(a)}_i=x_i(\C_f)-x_i(\C_0)$.

At the  macroscopic level, namely when only the exchanges of extensive quantities with the reservoirs are measured, there appears a symmetry for transient regimes where the system  is initially prepared in some equilibrium state with a fixed intensive parameter $F_i^0$ for each configuration observable $x_i$ and then is suddenly put into contact with reservoirs with thermodynamic parameters $F^{(a)}_i$'s that drive the system into a non-equilibrium state. 
The symmetry involves the excess exchange entropy variation $\Deltaexchexcessgen S$, defined as the difference between the exchange entropy variation under the non-equilibrium external constraints $F^{(a)}_i$'s, defined in \eqref{defgenDeltaexchS}, and the corresponding variation under the equilibrium conditions where for  all reservoirs  that exchange  quantity with index $i$  the  thermodynamic parameter has the same value
$F_i^0$, 
 \be
\Deltaexchexcessgen(\{{\cal X}^{(a)}_i\})\equiv\Deltaexch S (\{{\cal X}^{(a)}_i\})-\sum_i F_i^0\sum_{a\in R(i)} {\cal X}^{(a)}_i.
\ee 
When the system is prepared in the equilibrium state with probability distribution $\Probeq(\{F_i^0\})$ and put into contact with reservoirs with thermodynamic parameters $F^{(a)}_i$'s at the initial time of the measurements of exchanged quantities, the joint probability for the cumulative exchange quantities ${\cal X}^{(a)}_i$ obeys the symmetry relation at any finite time,
\be
\label{DFRJointProbChi}
\frac{\Prob_{\Probeq(\{F_i^0\})}\left(\{{\cal X}^{(a)}_i\}\right)}
{\Prob_{\Probeq(\{F_i^0\})}\left(\{-{\cal X}^{(a)}_i\}\right)}=e^{-\Deltaexchexcessgen S(\{{\cal X}^{(a)}_i\})}.
\ee
As a consequence, the excess exchange entropy variation $\Deltaexchexcessgen S$   obeys  the symmetry relation at any finite time, or ``detailed fluctuation relation'', 
\be
\label{DFRDeltaexchexcess0gen}
\frac{\Prob_{\Probeq(\{F_i^0\})}\left(\Deltaexchexcessgen S\right)}
{\Prob_{\Probeq(\{F_i^0\})}\left(-\Deltaexchexcessgen S\right)}=e^{-\Deltaexchexcessgen S}.
\ee
The latter relation itself entails the identity, or ``integral fluctuation relation'', 
\be
\Esp{e^{\Deltaexchexcessgen S}}_{\Probeq(\{F_i^0\})}=1.
\ee
We  notice that a relation similar to \eqref{DFRJointProbChi} has been derived from the  fluctuation relation for the entropy production  \cite{Seifert2005, Seifert2008} in a slightly different protocol where several reservoirs exchange both energy and matter with the system of interest and the system is prepared at equilibrium with one reservoir \cite{CuetaraEspositoImparato2014}.

\appendix

\section{Property of the  coarse grained dynamics}
\label{CoarseGrainedProperty}

In the present appendix we derive the property \eqref{NumbersEqual} valid over a period of the ergodic  deterministic microscopic dynamics $\Udisc$ when  $\Udisc$ respects both the conservation of $E_\text{dec}$ and the interaction pattern specified when the conservation law \eqref{conserE} was introduced.

Though we believe that a  general development could be pursued,  we
prefer to concentrate on a specific model at this point. 
The system  is made of two large
parts and a small one, which is reduced to  two Ising spins $\sigma_1, \sigma_2=\pm
1$, each one directly in contact with one of the large parts. So a configuration
$C$ can be written as $C=(C_1,\sigma_1,\sigma_2,C_2)$. We assume that, when the small part is isolated, its energy $\En(\sigma_1,\sigma_2)$ does not describe independent spins. Moreover, the microscopic dynamics $\Udisc$ conserves the energy $E_\text{dec}
(C)$ where the interactions between parts is neglected, namely $E_\text{dec}
(C)\equiv E_1(C_1)+E_2(C_2)+\En(\sigma_1,\sigma_2)$. 
 The remnant of interactions
between the parts is embodied in the following restrictions. For any $C$, $\Udisc(C)$
is obtained from $C$ by one of the following operations : 

- (I) a flip of spin $\sigma_1$ together with a change in $C_1$ and a possible change in 
$C_2$ such that $E_1(C_1)+\En(\sigma_1,\sigma_2)$ and $E_2(C_2)$ both remain
constant (i.e. the energy needed to flip the spin $\sigma_1$ entirely comes from or goes to 
large part $1$).

- (II) a flip of spin $\sigma_2$ together with a change in $C_2$ and a possible change in
$C_1$ such that $E_1(C_1)$ and $E_2(C_2)+\En(\sigma_1,\sigma_2)$ both remain
constant (i.e. the energy needed to flip the spin $\sigma_2$ entirely comes from or goes to
large part $2$).

- (III) a change in $C_1$ and/or $C_2$ but no flip of $\sigma_1$ or $\sigma_2$,
such that $E_1(C_1)$ and $E_2(C_2)$ both remain constant, as well as $\En(\sigma_1,\sigma_2)$.

In order to build an  effective mesoscopic dynamics we just keep track of $(E_1,\sigma_1,\sigma_2,E_2)$ as a function of time.  Therefore during  the  time evolution of a given configuration $C$ of the whole system we
concentrate only on  time steps at which a change of type (I) or (II) occurs, namely
when either spin $\sigma_1$ or spin $\sigma_2$ is flipped with known corresponding variations in $E_1$ and $E_2$. We do not follow
precisely the changes (III) that modifies the configurations of  large parts without changing the energy of any part (either $E_1$, $E_2$ or $\En(\sigma_1,\sigma_2)$). 
The possible changes are
\be
\label{typeI}
(E_1, \sigma_1,\sigma_2,  E_2)\rightarrow (E'_1, -\sigma_1, \sigma_2,E_2)\quad\textrm{with}\quad E'_1=E_1+\En(\sigma_1,\sigma_2)-\En(-\sigma_1,\sigma_2)
\quad\textrm{type (I)}\quad
\ee
and 
\be
\label{typeII}
(E_1, \sigma_1,\sigma_2,  E_2)\rightarrow (E_1, \sigma_1,-\sigma_2,E'_2)\quad\textrm{with}\quad E'_2=E_2+\En(\sigma_1,\sigma_2)-\En(\sigma_1,-\sigma_2)
\quad\textrm{type (II)}\quad
\ee

Starting from some initial configuration $(E_1^0,\sigma_1^0,\sigma_2^0,E_2^0)$,
some set of possible $(E_1,\sigma_1,\sigma_2,E_2)$ will be visited during
the time evolution over a period of $\Udisc$, and it is useful to view this set as the vertices of a
graph, whose edges connect two vertices if the system can jump from one to the other by
a single change of type (I) or (II). This graph can be chosen to be unoriented
because a transformation of type (I) or of type (II) is its own inverse.  Then
a trajectory over a period of the underlying
deterministic dynamics corresponds to a closed walk on this graph, which
summarizes the coarse graining due to the macroscopic description of the large
parts.  During a period of the microscopic dynamics the closed walk on  the
graph goes  through each edge a number of times.

The main observation is that, since energy $\En(\sigma_1,\sigma_2)$ does not
describe independent spins, the graph has the topology of a segment. Indeed, the
graph is connected by construction, but each vertex has at most two neighbors.
So the graph is either a segment or a circle. Let us suppose that it is a
circle. Then one can go from $(E_1^0,\sigma_1^0,\sigma_2^0,E_2^0)$ to itself by
visiting each edge exactly once, i.e. by an alternation of moves of type (I) and
(II).  After two steps both spins in the small part are flipped, so the total
length of the circle is a multiple of $4$.  But after $4$ steps a definite
amount of energy has been transferred between the two large parts, namely the
energy in the first large part has changed by
$\En(\sigma_1,\sigma_2)-\En(-\sigma_1,\sigma_2)+\En(-\sigma_1,-\sigma_2)-\En(\sigma_1,-\sigma_2)$
and a trivial computation shows that this cannot vanish unless
$\En(\sigma_1,\sigma_2)$ describes two independent spins, a possibility which
has been discarded. This excludes the circle topology.

Since the graph is a segment, any closed walk on the graph traverses a given
edge the same number of times in one direction and in the other one.  As a
consequence, during a period of the microscopic dynamics $\Udisc$, the motion
induced by $\Udisc$ on the graph is such that the number of transitions of type
(I) $(E_1,\sigma_1,\sigma_2,E_2)\rightarrow (E'_1,-\sigma_1,\sigma_2,E_2)$ is
equal to the number of their inverse transitions
$(E'_1,-\sigma_1,\sigma_2,E_2)\rightarrow(E_1,\sigma_1,\sigma_2,E_2)$, where the
relation between $E'_1$ and $E_1$ is given in \eqref{typeI}.  The same
considerations apply for transition rates associated with the flipping of
$\sigma_2$. The result is summarized in \eqref{NumbersEqual}.

For a more general discrete system, the mesoscopic time evolution can again be characterized by the time steps where the small system variables are flipped while the changes in the large parts are only macroscopically described by their net energies. An analogous graph can be constructed but it
can be much harder to analyze its topology, which is the crucial knowledge needed to
exploit the consequences of ergodicity. These consequences are most stringent
for a tree. There is no reason a priori why the graph should be a tree, but what
this implies, namely that ratios for the transition rates involving the small part and a
large part are given by ratios of energy level degeneracies in the large part, is
physically quite appealing.

\section{The Markovian approximation}
\label{Markovapprox}

This appendix is a short digression on mathematics. The
aim is to briefly recall a trick allowing to replace a fixed sequence of symbols
by a random one with ``analogous'' statistical properties, and then to combine this
trick with a coarse graining procedure.

\subsection{Discrete time stochastic approximation}
\label{DiscreteTMarkov}

Suppose $w \equiv x_1x_2\cdots x_N$ is a given finite sequence
of elements of a finite set $S$. It is convenient to assume that this sequence
is periodic, i.e. that $x_{N+1}\equiv x_1$. For $x\in S$, set $N_x\equiv \#\{i\in
[1,N], x_i=x\}$, i.e. $N_x$ is the number of occurrences of $x$ in the sequence
$w$. There is no loss of generality in assuming that $N_{x}\neq 0$ for
every $x\in S$, should it lead to consider a smaller $S$. For $x,x' \in S$ set
$N_{xx'}\equiv \#\{i\in [1,N], x_i=x,x_{i+1}=x'\}$, i.e. $N_{xx'}$ is the number of
occurrences of the pattern $xx'$ in the sequence $w$. Notice that we accept
that $x=x'$ in this definition. Of course, we could look at the occurrence of more
general patterns. By definition, we have $\sum_{x'\in S} N_{xx'}=\sum_{x'\in
  S}N_{x'x} =N_x$. The result we want to recall is the following.

There is a single time-homogeneous irreducible Markov matrix on $S$ that fulfills the following two requirements. First, 
in the stationary state of the corresponding discrete-time stochastic evolution,
the probability $\Probst(x',i+1;x,i)$ that a sample $\widehat{x}$ takes the
value $x$ at time $i$ and the value $x'$ at time $i+1$\footnote{In the whole paper we use the convention that the
  evolution (here given by the joint probability $\Probst(x',i+1;x,i)$ or the transition matrix $T$) is written from the right
  to the left as in quantum mechanics.} is equal to the
frequency of the pattern $xx'$ in the sequence $w$, namely 
\be
\label{propequiv2}
\Probst(x',i+1;x,i)=\frac{N_{xx'}}{N}.
\ee
Second, the stationary probability that the random variable $\widehat{x}$
takes the value $x$ at any time $i$ is equal to the frequency of $x$ in the
sequence $w$, namely
\be
\label{propequiv1}
\Probst(x)=\frac{N_x}{N}.
\ee
In fact \eqref{propequiv1} is  a consequence of \eqref{propequiv2}, because of
the relations $\Probst(x)=\sum_{x'\in S}\Probst(x',i+1;x,i)$ and
$\sum_{x'\in S} N_{xx'}=N_x$.  To say it in words, there is a unique
time-homogeneous irreducible Markov chain whose stationary statistics for
patterns of length $1$ or $2$ is the same as the corresponding statistics in
$w$.

The proof is elementary. By the Markov property, the Markov matrix element from
$x$ to $x'$, denoted by $T(x' \leftarrow x)$, must satisfy
the relation $\Probst(x',i+1;x,i)= T(x' \leftarrow x)\Probst(x)$, so the only
candidate is 
\be
\label{propequiv3}
T(x' \leftarrow x)=\frac{N_{xx'}}{N_x}.
\ee
Conversely, the corresponding
matrix is obviously a Markov matrix ($\sum_{x'\in S}T(x' \leftarrow x)=1$ since
$\sum_{x'\in S}N_{xx'}=N_x$), and it is irreducible, because, as $w$ contains all
elements of $S$, transitions within $w$ allow to go from every element of $S$ to
every other one. Checking that for this Markov matrix the stationary measure, namely
the solution of $\sum_{x\in S}T(x' \leftarrow x) \Probst(x) =\Probst(x')$, is
given by \eqref{propequiv1} boils down to the identity $\sum_{x\in S}
N_{xx'}=N_{x'}$ recalled above, and then the two-point property \eqref{propequiv2}
follows. This finishes the proof.

In the very specific case where $x_1,x_2,\cdots,x_N$ are all distinct, i.e. $|S|$, the cardinal of $S$, is equal to $N$, $N_x=1$
for all $x\in S$ and $N_{xx'}=\delta_{x',x_{i+1}}$ where $i$ is such that $x_i=x$.
Then the only randomness lies in the choice of the initial distribution, and
each trajectory of the Markov chain reproduces $w$ up to a  translation of  all indices. If $N$ is
large and $|S|$ is $\sim N$, slightly weaker but analogous conclusions survive.
A more interesting case is when $|S| \ll N$ by many orders of magnitude as discussed in next subsection.

This trick has been used for instance to write a random text ``the Shakespeare
way'' by computing the statistics of sequences of two words in one of his
books.

By definition, the Markovian approximation reproduces the statistics of the
original sequence only for length $1$ and length $2$ patterns. It is a delicate
issue to decide whether or not it also does a reasonnable job for other
patterns. For instance, the random Shakespeare book certainly looks
queer. Various physical but heuristic arguments suggest that, for the kind of
sequences $w$ relevant for this work, the Markov approximation is quite
good, but we shall not embark on that.

\subsection{Continuous time approximation}
\label{ContinousTimeApproximation}

By the very same argument, there is a  time-homogeneous irreducible Markov
transition matrix on $S$ (unique up to a time scale  $\tau$) such that, in the
stationary state of the corresponding continuous-time stochastic evolution, the
expected number of transitions from $x$ to $x'\neq x$ per unit time is
$N_{xx'}/\tau$. The formula one finds for the transition rate is 
\be
\label{Wxxprim} 
W(x'\leftarrow x)= \frac{N_{xx'}}{\tau N_x} \text{ for }x \neq x'.
\ee

The continuous-time approximation becomes more natural in the case when $|S|$,
the cardinal of $S$, is such that $|S|\ll N$, while for all $x$ $N_{xx}\sim N_x$ and for all $x \neq x'$ $N_{xx'} \ll N_x$, which means that transitions are rare and most of the time
$x$ follows $x$ in the sequence $w$.  Again, one can argue that for the kind of
sequences $w$ relevant for this work, this is guaranteed by physics. Then taking
$\tau$ (in some macroscopic time unit) of the order of the largest value of the
$N_{xx'}/N_x$, $x \neq x'$, one ends with a Markov transition matrix with
elements of order unity (in some macroscopic inverse time unit), and $t_i=\tau
i$ can be taken as the physical macroscopic time.

To summarize, in this work, we shall systematically associate to certain
sequences $w$ a continuous-time Markov process and exploit properties of $w$ to
constraint the structure of $W(x' \leftarrow x)$.

\vspace{.3cm}

A natural application of the above ideas is to the case where the sequence $w$ arises from some coarse graining procedure. One starts from a
sequence $\omega=\xi_1\xi_2\cdots \xi_N$ where $\xi_1,\xi_2,\cdots ,\xi_N$
belong to a set $\Sigma$ that is so large that $\omega$ cannot be stored, and
that only some of its features can be kept. Say we partition $\Sigma\equiv \cup_{x
  \in S} \Sigma_x$ where $S$ is of reasonable size.  Then all we keep of
$\omega$ is $w=x_1x_2\cdots x_N$ where, for $i\in [1,N]$, $x_i$ is substituted
for $\xi_i$ when $\xi_i\in \Sigma_{x_i}$. In the applications we
have in mind, $\Sigma$ is an $N$-element set, i.e. all terms in the sequence
$\omega$ are distinct. In that case, even if $\omega$ is constructed in a
perfectly deterministic way, by saying who follows who in the sequence, such a
description is unavailable on $w$, and $w$ may well look quite random, so the
Markov chain approximation is worth a try. In fact, $|S| \ll N$ and we shall
take as a physical input that transitions are rare, so that the (continuous-time)
Markov process is an excellent approximation to the (discrete time) Markov chain.

\section{Microcanonical detailed balance and time reversal invariance in Hamiltonian dynamics}
\label{TimeReversal}

A microcanonical detailed balance similar to \eqref{ratioWmc} can be derived in another context where the  discrete microscopic  variables of the whole system evolving under an ergodic energy-conserving map are replaced by continuous  microscopic coordinates $\xi$ in phase space evolving under  a deterministic dynamics whose Hamiltonian  $H[\xi]$ is an even function of momenta, and where  the role of the mesoscopic configurations  $(\C,\underline{E})$ is played by some continuous mesoscopic variables $x[\xi]$ that are even functions of momenta (with the same notations as in subsubsection \ref{DefinitionMarkovApprox}).  

A summary of the argument is the following. From the viewpoint of the statistical ensemble theory, the initial position  of the whole system in phase space is not known, and it is assumed to be     uniformly randomly distributed in the energy shell $E=H[\xi]$, where $E$ is the value of the energy of the full system at the initial time $t=0$.  In other words, the initial probability distribution 
$\Prob_0(\xi)$ is such that $d\xi\,\Prob_0(\xi)\equiv d\xi \,\delta\left(H[\xi]-E\right)/\int d\xi'\, \delta\left(H[\xi']-E\right)$. Since both the  Liouville measure $d\xi$ and the energy $H[\xi]$ are conserved when  the system  evolves  in phase space from an initial position  $\xi$ at time $t=0$ to a  position $f_t(\xi)$ at time $t$,  $\Prob_0(\xi)d\xi$ is conserved by  the Hamiltonian dynamics. As a consequence
 the  probability that at time $t$ the set of mesoscopic variable $x$ takes the value $x_1$,  defined as
$\Prob_{\Prob_0}(x_1, t)\equiv\int d\xi \,\Prob_0(\xi)\,
\delta\left(x[f_t(\xi)]-x_1\right)$
is in fact independent of time, and it coincides with  the microcanonical  distribution $\Probmc(x_1)\equiv\int d\xi \,\Prob_0(\xi)\,
\delta\left(x[\xi]-x_1\right)$.

Moreover,  since the Hamiltonian is assumed to be an even function of momenta, the microscopic dynamics is invariant under time reversal : $[Rf_tRf_t](\xi)=\xi$,
 where    $R$ is the operator that changes every momentum into its opposite.
 As a consequence, if the  mesoscopic variables $x[\xi]$ are even functions of momenta, 
the time-displaced joint probability that the set  $x$ takes the value  $x_1$ at time $t=0$ and the value $x_2$ at time $t$, defined as
$\Prob_{\Prob_0}(x_2,t;x_1,0)=\int d\xi \,\Prob_0(\xi)\,
\delta\left(x[f_t(\xi)]-x_2\right) 
\delta\left(x[\xi]-x_1\right)$, 
 is invariant under the exchange of the times at which $x_1$ and $x_2$ occur, 
\be
\label{ProbJointHBis}
\Prob_{\Prob_0}(x_2,t;x_1,0)=\Prob_{\Prob_0}(x_1,t;x_2,0).
\ee
Morevover the displaced joint probability is invariant under time translation.

Then the approximation scheme by a Markov process for the time-displaced joint probability of the continuous mesoscopic variables  $x$ is the same as that we have used for the two-body ergodic average of the discrete mesoscopic configurations $(\C,\underline{E})$ in subsubsection \ref{DefinitionMarkovApprox}.
One assumes that the evolution of  the mesoscopic variable 
$x$ described by $\Prob_{\Prob_0}(x_2,t;x_1,0)$  can be approximated  by an homogeneous  Markovian stochastic process whose stationary distribution $\Probst(x)$ is the time-independent probability $\Prob_{\Prob_0}(x)$  and whose  transition rates, denoted by $W(x'\leftarrow x)$, are determined by  the identification 
$\Prob_{\Prob_0}(x', dt;x,0)\equiv W(x'\leftarrow x) \Prob_{\Prob_0}(x)\times dt$.
Then the time reversal symmetry property \eqref{ProbJointHBis} for the joint probability of $x$ and $x'$ at different times and the fact that $ \Prob_{\Prob_0}(x)$ coincides with the microcanonical distribution 
$ \Probmc(x)$ lead  to the microcanonical detailed balance
\be
 W(x'\leftarrow x) \Probmc(x)= W(x\leftarrow x') \Probmc( x').
\ee

\bibliographystyle{unsrt}

\end{document}